 \documentclass[prl,amsmath,amssymb,twocolumn, showpacs, superscriptaddress,10pt]{revtex4-1}

\usepackage{amsmath}
\usepackage{hyperref}
\usepackage{graphicx}
\usepackage{amsfonts}
\usepackage{amsthm}
\usepackage{cases}
\usepackage{bm}

\usepackage[normalem]{ulem}

\usepackage{color}
\definecolor{Blue}{rgb}{0.00, 0.00, 1.00}
\definecolor{Red}{rgb}{1.00, 0.00, 0.00}
\definecolor{Green}{rgb}{0.00, 0.70, 0.00}

\hypersetup{
    colorlinks=true,       
    linkcolor=red,          
    citecolor=blue,        
    filecolor=magenta,      
    urlcolor=cyan           
}

\newcommand{\nn}{\nonumber}
\newcommand{\be}{\begin{equation}}
\newcommand{\ee}{\end{equation}}
\newcommand{\bea}{\begin{eqnarray}}
\newcommand{\eea}{\end{eqnarray}}

\newcommand{\beq}{\begin{equation}}
\newcommand{\eeq}{\end{equation}}
\newcommand{\beqn}{\begin{eqnarray}}
\newcommand{\eeqn}{\end{eqnarray}}

\begin{document}

\title{
Ranked diffusion, delta Bose gas and Burgers equation}

\author{Pierre Le Doussal}

\affiliation{Laboratoire de Physique de l'\'Ecole Normale Sup\'erieure, CNRS, ENS and PSL University, Sorbonne Universit\'e, Universit\'e de Paris, 75005 Paris, France}

\date{\today}

\begin{abstract}
We study the diffusion of $N$ particles in one dimension interacting via a drift proportional to their rank.
In the attractive case (self-gravitating gas) a mapping to the Lieb Liniger quantum model allows to
obtain stationary time correlations, return probabilities and the decay rate to the stationary state. 
The rank field obeys a Burgers equation, which we analyze. 
It  allows to obtain the stationary density at large $N$ 
in an external potential $V(x)$ (in the repulsive case). In the attractive case the decay rate 
to the steady state is found to depend on the initial condition if its spatial decay is slow enough.
Coulomb gas methods allow to study the final equilibrium at large $N$. 
\end{abstract}



\maketitle

Interacting ranked diffusion, i.e. the diffusion of $N$ particles in 1D under a drift which depends only on their rank
were used to model financial or economic data \cite{Banner}. Pal and Pitman \cite{Pitman} studied the case where each particle feels a drift $\delta_j$ where $j$ is the rank of the particle ($j=1$ is the leftmost one etc..).
They showed that the particle spacings converge to independent exponential variables with
rates $2 \alpha_j$, where $\alpha_j = \sum_{i=1}^j (\delta_i - \bar \delta)$ and 
$\bar \delta$ is the average drift. It holds provided $\alpha_k>0$ for all $1 \leq k \leq N-1$, i.e. for attractive interactions.
They showed connections to reflected Brownian motions in wedges with
drifts, generalized by O'Connell and Ortmann \cite{OConnell}. The large $N$ limit was studied in \cite{Jourdain2000,JourdainChaosProp2008,JourdainReygner2013,Reygner2015,Pal}
and shown to be related to a non-linear diffusion process, an example of a more general phenomenon known as propagation of chaos \cite{Kac,McKean,Sznitman1,Sznitman2,Calderoni}. 

Since the Coulomb interaction in 1D is linear in the distance, these models in their stationary state
are related to the statistical mechanics of the self-gravitating 1D gas \cite{Rybicki,Kumar2017} 
(attractive case), or of the 1D Coulomb gas (CG) (repulsive case) in presence of a background charge or in a finite box, 
also called Jellium \cite{Jellium,SatyaJellium1,SatyaJellium2,SatyaJellium3}. 

In this paper we examine the dynamics of this model, in the light of two exact mappings. (i) The first one, useful 
for attractive interactions, is to the Lieb-Liniger delta Bose gas model. It allows to obtain the relaxation spectrum for
a class of initial conditions with fast enough spatial decay. It also leads some return probabilities, and time dependent correlation functions in the stationary state. (ii) The second is to a Burgers equation with noise, 
where the noise is subdominant in the large $N$ limit. 
We use it to determine the stationary state, in the repulsive case in presence of an external potential,
as well as the relaxation rate in the attractive case, for initial conditions with slow spatial decay.
These lead to a decay rate which depends on the initial condition. 
Finally we discuss the connection to the Coulomb gas, which allows to determine the
true final equilibrium state. 

We consider $N$ particles on the real line at positions $x_i(t)$ evolving according to the Langevin equation
\bea \label{langevin1}
\frac{dx_i}{dt} = \bar c \sum_{j=1}^N {\rm sgn}(x_j-x_i) - V'(x_i) + \sqrt{2 T} \xi_i(t) 
\eea 
where $\xi_i(t)$ are unit independent white noises, $T$ the temperature, and ${\rm sgn}(0)=0$.
Here $V(x)$ is an external potential, seen by all walkers.
The particles will cross, and we denote $x_{(j)}(t)$ the ordered sequence
of their positions at time $t$, i.e. $x_{(1)}(t) \leq x_{(2)}(t) \dots \leq x_{(N)}(t)$. 
The ordered particle $x_{(j)}$ then feels the permanent drift 
$\delta_j=(N+1- 2 j) \bar c$. The case $\bar c>0$ corresponds to attractive
interactions and the system (apart from its center of mass) 
reaches a stationary state even if $V(x)=0$. 
In the case of repulsive interactions, $\bar c<0$, it is useful to add
a confining potential, such as an harmonic well, $V(x)=\frac{1}{2} \mu x^2$,
or a linear trap $V(x)=\mu |x|$, with $\mu>0$.

The probability distribution function (PDF), $P(\vec x,t)$, of a given configuration,
${\vec x}= \{ x_i(t) \}_{i=1,\dots,N}$, satisfies the 
Fokker-Planck (FP) equation 
\bea \label{FP0} 
\partial_t P &=& - {\cal H}_{\rm FP} P \\
&=&  \sum_i [ T \partial_{x_i}^2 
+ \partial_{x_i} (V'(x_i) + \bar c \sum_j {\rm sgn}(x_i-x_j) ) ] P \nn
\eea 
It admits a zero current stationary PDF
\bea \label{PstatV} 
&& P_{\rm stat}(\vec x) = \frac{1}{Z_N} e^{- \frac{\bar c}{2 T} \sum_{i,j=1}^N |x_i-x_j| 
- \frac{1}{T} \sum_{i=1}^N V(x_i) } 
\eea
which is normalizable on the line when $V(x)$ is a confining potential. 
In the attractive case, $\bar c >0$, and setting $V(x)=0$,
one can rewrite \cite{footnotecdm}
\be \label{Pstat} 
P_{\rm stat}(\vec x) \propto e^{ \frac{\bar c}{T} \sum_{j=1}^N (N+1-2j) x_{(j)} } 
= e^{- \sum_{j=1}^{N-1} \alpha_j (x_{(j+1)} - x_{(j)} )}
\ee 
with $\alpha_j=\sum_{i=1}^j \delta_i/T$ in agreement with \cite{Pitman}
(which uses $T=1/2$) since here $\bar \delta=0$. For $\bar c <0$ the particles repel each others and one needs 
a confining potential. In this case $P_{\rm stat}(\vec x)$ can be interpreted as the equilibrium Gibbs measure of a 1D CG, e.g. as recently studied in  \cite{SatyaJellium1,SatyaJellium2} in the
case of the harmonic well.

{\it Mapping to the delta Bose gas}. The ranked diffusion (RD) \eqref{langevin1}
with $V(x)=0$ can be mapped to the Lieb-Liniger (LL) model of delta interacting quantum particles.
Defining $\Psi_0(\vec x)=C P_{\rm stat}(\vec x)^{1/2}$ from \eqref{Pstat} \cite{footnoteC} the FP operator \eqref{FP0} 
relates to a Schr\"odinger operator ${\cal H}_s$ via
\be
{\cal H}_{\rm FP}  = \Psi_0 {\cal H}'_{s} \Psi_0^{-1} 
\ee 
where ${\cal H}'_{s} = {\cal H}_{s} - E_0$ and (setting $T=1$ in this section)
\be \label{LL} 
{\cal H}_s =  - \sum_i \partial_{x_i}^2  - 2 \bar c \sum_{1 \leq i < j \leq N} \delta(x_i-x_j) 
\ee
is the Hamiltonian of the LL model \cite{LL}, with $\bar c=-c$ in standard notations.
This mapping is useful in the attractive case, $\bar c>0$, since the the bound state $\Psi_0$ (also called $N$-string)
is the ground state of ${\cal H}_s$, with energy $E_0=- \frac{1}{4} \sum_{i=1}^N \delta_i^2 = - \frac{\bar c^2}{12} (N^3-N)$ and zero center of mass momentum. By contrast in the the repulsive case $\bar c<0$, it is not the ground state of the LL model \cite{footnote2} (for some $V(x)$ it can be related to a quantum model, but with additional interactions 
\cite{SM}).  
Since for $V(x)=0$ the LL model is integrable, so is the dynamics of the RD system. 
The Green's functions 
$G_{\rm FP/s}=\langle x | e^{- t {\cal H}_{\rm FP/s}} |y \rangle$ are related via
\be
G_{\rm FP}(\vec x, \vec y,t) = \frac{\Psi_0(\vec x)}{\Psi_0(\vec y)} G_{s}(\vec x, \vec y,t) e^{E_0 t} 
\ee
and the time dependent PDF of the RD system is obtained from the imaginary time dynamics of the LL
model as 
\bea \label{Pt} 
&& P(\vec x,t)  = \Psi_0(\vec x)  {\cal Z}_N(\vec x,t) e^{E_0 t} \\
&& {\cal Z}_N(\vec x,t)  = 
 \sum_\lambda  \frac{\langle \Psi_\lambda| \Psi(0) \rangle}{||\Psi_\lambda||^2} \Psi_\lambda(\vec x)  e^{- E_\lambda t} \nn
\eea
where the $E_\lambda$ are $\Psi_\lambda$ are the eigenenergies and the (unnormalized) 
eigenstates of ${\cal H}_s$, given by the Bethe ansatz, and
$\langle \vec x| \Psi(0) \rangle=\Psi(\vec x,t=0)=P(\vec x,0)/\Psi_0(\vec x)$
is the initial condition.
For a general initial condition (IC), 
eigenstates of ${\cal H}_s$ with arbitrary symmetry contribute
(the model being called Gaudin-Yang \cite{GaudinYang}).
Choosing a symmetric IC for the RD problem allows to restrict to bosonic
eigenstates. One then deduces the relaxation spectrum 
of the FP operator \eqref{FP0} of the RD system as $E_\lambda - E_0$.
The eigenstates \cite{m-65} are indexed by the partitions of $N$, i.e. the sets of integers $m_i \geq 1$ such
that $\sum_{i=1}^{n_s} m_i=N$, together with wavevectors $k_i \in \mathbb{R}$
\be
E_\lambda = - \frac{\bar c^2}{12} \sum_{i=1}^{n_s} (m_i^3-m_i) + \sum_{i=1}^{n_s}  m_i k_i^2 
\ee
interpreted as the energy of $1 \leq n_s \leq N$ bound states, each with $m_i$ particles
and center of mass momentum $m_i k_i$. The ground state $\Psi_0$ has
$n_s=1$, $m_1=N$ and $k_1=0$. We call "ground state manifold" the space of
superpositions of $N$-string eigenstates with momenta $k_1=k$, 
$\Psi_k(\vec x)=e^{i k \sum_{i=1}^N x_i} \Psi_0(\vec x)$. It is invariant by the
dynamics. These superpositions describe the RD system in its
stationary state in terms of relative coordinates, together with the free diffusion of the center of mass 
$\bar x=\frac{1}{N}\sum_i x_i$, i.e. $P(\vec x,t) \propto \Psi_0(\vec x)^2 p_t(\bar x)$ with
$p_t(x) = e^{- (t/N) \partial^2_x} p_0(x)$. 
In terms of relative coordinates, the lowest excited state has $n_s=2$, $m_1=N-1$, $m_2=1$,
i.e. one particle evaporating from the ground state. The convergence to the ground
state manifold is thus exponential as $\sim e^{- \gamma_N t}$,
with relaxation rate
\be \label{relax1} 
\gamma_N = \frac{\bar c^2}{4} N(N-1) 
\ee 
There is a catch however. Eq. \eqref{Pt} holds only insofar the overlaps
$\langle \Psi_\lambda| \Psi(0) \rangle$ exist, i.e. are given by convergent
integrals. This is the case for IC such that $P(\vec x,0)$ decays sufficiently fast
(typically as $\sim \Psi_0(\vec x)$ or faster) at large $\vec x$. Then the decay rate is 
given by \eqref{relax1} and is independent of further details of the IC. 
Consider e.g. the case $N=2$ which is simple to solve by other methods,
and $P(\vec x,0) \propto e^{- a |x_1-x_2|}$. One checks \cite{SM} that
for $a > \bar c/2$ Eq. \eqref{relax1} holds, while for $a< \bar c/2$,
one has $\gamma_2 = 2 a(\bar c -a)$, i.e. 
the decay is slower and its rate depends on the IC. Although it seems possible
to obtain that decay from an analytic continuation of \eqref{Pt}, we do not pursue it here.
Below we obtain this decay at large $N$ using the Burgers equation,
confirming the above predictions.

Sums as ${\cal Z}_N(\vec x,t)$ in \eqref{Pt} have been studied recently in the
context of the KPZ equation/directed polymer problem, as recalled in 
\cite{SM}. For symmetric IC
one has ${\cal Z}_N(\vec x,t) = \mathbb{E}[e^{\sum_{i=1}^N h(x_i,t)}]$, i.e. an average 
over the noise of the solution $h(x,t)$ of the KPZ equation \cite{KPZ} such that 
${\cal Z}_N(\vec x,t=0) =P(\vec x,0)/\Psi_0(\vec x)$. The so-called
droplet IC thus corresponds to $P(\vec x, t=0)= \prod_i \delta(x_i)$. 
Using known results
\cite{PLDdroplet,we-flat,we-flatlong,flat-shorttime}, we display in \cite{SM} the exact formula for the return probability $P(\vec 0,t)$
in that case, as well as for the IC $P(\vec x,t=0) \propto \Psi_0(\vec x)$, which
corresponds to the flat IC for the KPZ equation. Note that for symmetric IC 
the propagator $G_{s}(\vec x, \vec y,t)$ is known explicitly \cite{ProhlacSpohnPropagator}. 

Finally, one can show that correlation functions in the stationary state of the RD system,
such as the two time correlation of the particle density, can be obtained from their analog in the 
attractive LL model. Using the results of Caux and Calabrese 
\cite{CalabreseCauxBosonsPRL,CalabreseCauxBosonsLong}
we display an exact formula for
this correlation function in \cite{SM}. 


{\it Mapping to a stochastic Burgers equation}. We now turn to a completely
different method which applies for any sign of $\bar c=-c$ and in presence of an external
potential $V(x)$. Let us define the empirical density 
$\rho$ normalized to unity, 
$\rho(x,t) = \frac{1}{N} \tilde \rho(x,t)$ where $\tilde \rho(x,t)=\sum_i \delta(x-x_i(t))$.
Using the Dean-Kawasaki method \cite{Dean}, one obtains a
closed stochastic equation for its evolution, which is exact for any $N$
\cite{footnote1}
\bea \label{eqrho1}
&& \partial_t \rho(x,t)  =   T \partial_x^2 \rho(x,t)  + \frac{1}{\sqrt{N}} \partial_x [ \sqrt{ 2 T \rho(x,t) } \eta(x,t) ] \\
&& +  \partial_x [ V'(x) \rho(x,t) +
N \bar c \rho(x,t) \int dy \rho(y,t) {\rm sgn}(x-y) )] \nonumber 
\eea 
where $\eta(x,t)$ is a normalized spatial white noise. Now define the {\it rank field} $r(x,t)$ through
\be \label{defrank} 
\rho(x,t) = \partial_x r(x,t) ~,~ r(x,t) = \int^x_{-\infty} dx' \rho(x',t) - \frac{1}{2}
\ee 
which increases monotonically from $-1/2$ at $x=-\infty$ to $+1/2$ at $x=+\infty$.
Substituting $\rho(x,t)$ in \eqref{eqrho1}, we use that by integration
by part $\int dy \rho(y,t) {\rm sgn}(x-y) = 2 r(x,t)$, since 
$ [r(y,t) {\rm sgn}(x-y)]^{y=+\infty}_{y=-\infty} =0$. Integrating once
with respect to $x$ we obtain
\bea \label{eqr} 
 \partial_t r(x,t) &=& T \partial_x^2 r(x,t) + \frac{1}{\sqrt{N}}  \sqrt{ 2 T \partial_x r(x,t) } \eta(x,t) \nonumber \\
&+& 2 N \bar c r(x,t) \partial_x r(x,t) 
+ V'(x) \partial_x r(x,t)  
\eea 
since the integration constant vanishes from the boundary conditions for $r(x,t)$ at
$x=\pm \infty$. If $V(x)=0$ this is the Burgers equation with some multiplicative noise. Note that the
function $r(x,t)$ is constrained to be increasing in $x$ (positive density). We now consider
the large $N$ limit in two stages. \\

{\it Large $N$ at fixed $\bar c=-c$}. If we scale $V(x)=N \tilde V(x)$ and rescale time as
$t=\tau/N$, we can rewrite \eqref{eqr} as
\be
\partial_\tau r =  2  \bar c ~ r \partial_x r + \tilde V'(x) \partial_x r + \frac{1}{N} \partial_x^2 r 
+ \frac{1}{N}  \sqrt{ 2 \partial_x r } ~ \tilde \eta(x,\tau) 
\ee
still valid for any $N$, where $\tilde \eta$ is another unit white noise
(obtained from $\eta$ by the time change). In the limit $N \to +\infty$ we find
that $r(x,\tau)$ satisfies (with $c=-\bar c$)
\bea \label{1storder} 
\partial_\tau r = (\tilde V'(x) - 2 c  ~ r ) \partial_x r 
\eea 
We now consider separately the repulsive and attractive cases.

In the repulsive case, $c=-\bar c>0$, the only singularities can be plateaus
where $\partial_x r=0$, i.e. regions empty of particles. 
Setting $\partial_\tau r=0$ in \eqref{1storder} we see that around a given $x$ a 
stationary solution is either constant, $\partial_x r=0$, or equal to $r=\frac{1}{2} \tilde V'(x)/c$,
which is acceptable only if $\tilde V''(x)/c \geq 0$. The simplest
case is when $\tilde V(x)$ is convex. In that case the stationary solution is unique and given by
\be \label{aga0} 
r_{\rm stat} (x)= \frac{\tilde V'(x)}{2 c}  ~,~   \rho_{\rm stat} (x)=\frac{\tilde V''(x)}{2 c}  ~,~
x_e^- < x < x_e^+ 
\ee 
and $r_{\rm stat} (x)=\frac{1}{2} {\rm sgn}(x)$ elsewhere.
The support of the density is thus an interval and $x_e^\pm$ are the two edges, 
given by the roots of $\tilde V'(x_e^\pm)=\pm c$.
The density thus generically has a jump at these edges. This assumes that
the potential is sufficiently confining so that the roots $x_e^\pm$ exist. In the
case when $\tilde V'(+\infty)<c$ and $\tilde V'(-\infty)>-c$ or both, the edges are pushed to
infinity and a finite fraction of the particles are expelled to $\pm \infty$. 

For double (or multiple) well types potentials the situation is more involved, since there
are regions with $\tilde V''(x) <0$ where \eqref{aga0} cannot hold. 
There are thus families of stationary states, which are empty in these regions, leading
to multiple intervals support for the density \cite{SM}. 

What is the relaxational dynamics toward stationarity? From \eqref{1storder} we see that $r(x,\tau)$ increases with $\tau$ in regions such that $r < \tilde V'(x)/2c$ and decreases if $r>\tilde V'(x)/2c$. In the simplest case it leads to the
convergence to the unique stationary solution, Eq. \eqref{aga0}. For the double well, the dynamics leads the
system to one member of the family of stationary states, which depends on the initial condition $r_0(x)=r(x,0)$: it is easily found by noticing that at any point $x=a$ such that $r_0(a)=\tilde V'(a)/2 c$, one has $\partial_\tau r(a,\tau)=0$, hence
$r(a,\tau)=r_0(a)$. If at this point $\tilde V''(a)<0$, $r(x,\tau)$ must develop a plateau around this point in the large time limit, which determines uniquely the stationary state within the family, for details see \cite{SM}. 

The general solution of \eqref{1storder} is obtained by considering $\tau(x,r)$ and
using $\frac{\partial_x r}{\partial_\tau r}= - \partial_x \tau$. We obtain 
$(\tilde V'(x)-2 c r)\partial_x \tau=-1$. Let us denote $x_0(r)$ the inverse
function of the initial condition $r_0(x)=r(x,0)$, i.e. $r_0(x_0(r))=r$.
The general solution is then
\be
\tau=\int_{x_0(r)}^x \frac{dy}{2 c r - \tilde V'(y)} 
\ee 
For the harmonic well, $\tilde V'(y)=\mu_0 y$, with all particles starting at $x=y$, i.e.
$x_0(r)=y$, one finds that the density is uniform 
\be
r(x,\tau) = \frac{\mu_0}{2 c} \frac{x-y e^{-\mu_0 \tau}}{1- e^{- \mu_0 \tau} }
~,~ \rho(x,\tau)= \frac{\mu_0}{2 c(1- e^{- \mu_0 \tau})}
\ee
in the time dependent interval $x \in [x_e^-(\tau),x_e^+(\tau)]$ with $x_e^\pm(\tau) = \pm \frac{c}{\mu_0} (1- e^{- \mu_0 \tau}) + y e^{-\mu_0 \tau}$, and zero outside. It converges to the stationary state
$\rho_{\rm stat} (x) = \frac{\mu_0}{2 c}  \theta(\frac{c}{\mu_0} -|x|)$ as found in the Jellium studies.

For $\tilde V(x)=0$ one recovers the perturbative solution of the inviscid Burgers equation 
\be
2 c r \tau = x-x_0(r) \quad , \quad r = r_0(x - 2 c \, r \tau) 
\ee
equivalently $r(x,\tau) = r_0(w(x,\tau))$ where
\bea
w + 2 c ~ r_0(w) \tau = x  \quad  \Leftrightarrow \quad w=w(x,\tau) 
\eea 
It is valid as long as the map is invertible, i.e. $1 + 2 c \tau \rho_0(x) >0$
for all $x$, which always holds for $c>0$. An example is the square density 
initial condition, $\ell>0$
\bea \label{solurep0}
\rho(x,\tau) = \frac{1}{2 (\ell+ c \tau)} \theta(\ell + c \tau-|x|)
\eea 
which shows that the repulsive gas expands linearly in time.

Let us consider now the attractive case, $c=-\bar c<0$, and focus on $V(x)=0$.
Eq. \eqref{solurep0} is still valid, but now the gas contracts ballistically
and at time $\tau=\ell/ \bar c$ the density becomes a delta peak at $x=0$
containing all the particles, i.e. $r(x,\tau)$ develops a shock (a step). 
For more general initial conditions, the 
perturbative solution fails and a shock appears at time
\bea
\tau=\tau_s = \frac{1}{2 \bar c \max_x \rho(x,0)}
\eea 
and position $x=x_s= {\rm argmax} \rho(x,0)$. To describe the dynamics
with shocks one recalls that \eqref{1storder} originates from \eqref{eqr}.
The proper solution is then (see
next paragraph), 
$r(x,\tau) = \frac{w(x,\tau)-x}{2 \tau \bar c} = r_0(w(x,\tau))$
where 
\be 
w(x,\tau) := {\rm argmin}_{w \in \mathbb{R}} [ \frac{(w-x)^2}{4 \tau} - \bar c \int_0^w dx' r_0(x') ] \label{minimiz0} 
\ee
which recovers the perturbative solution when there is a single minimum in \eqref{minimiz0},
e.g. for $\bar c=-c<0$. For $\bar c>0$ \eqref{minimiz0} leads at intermediate time to one or several shocks, 
containing finite fractions of the total number of particles, which merge into a single one with unit fraction 
at some larger time, see \cite{SM} for details.


{\it Large $N$ with $\gamma=c N$, $\bar \gamma=\bar c N$ fixed}. 
Going back to \eqref{eqr} one can still neglect the noise but one must keep the diffusion term,
leading to 
\be \label{eqrho22} 
\partial_t r =  T \partial_x^2 r  + 2 \bar \gamma ~ r \partial_x r + V'(x) \partial_x r 
\ee 
Although interactions are different, this bears analogy with the studies 
\cite{BouchaudGuionnet} of matrix Brownian motion with index
$\beta \sim 1/N$, and here too because of the diffusion  
the stationary density does not vanish anywhere.

Consider first $V(x)=0$. Defining $Z$ a solution of
the heat equation, $\partial_t Z = T \partial^2_x Z$, and 
$r= \frac{T}{\bar \gamma} \partial_x \log Z$, leads to the Cole-Hopf solution (valid for any sign of $\bar \gamma$)
\bea \label{solu}
r(x,t) 
= \frac{\int \frac{dw}{\sqrt{4 \pi T t}} \frac{w-x}{2 \bar \gamma t} e^{ - \frac{(w-x)^2}{4 T t} + \frac{\bar \gamma}{T} 
 \int_0^w dx' r_0(x') } }{
\int \frac{dw}{\sqrt{4 \pi T t}} e^{ - \frac{(w-x)^2}{4 T t} + \frac{\bar \gamma}{T}  \int_0^w dx' r_0(x') }}
\eea 
where $r_0(x)=r(x,0)$ is the initial condition.
If one sets $t = \tau/N$ and $\bar \gamma = N \bar c$ the argument of the exponential is uniformly 
of $O(N)$ at large $N$ and one recovers the solution of the inviscid Burgers equation
\eqref{minimiz0}. 

In the attractive case, $\bar \gamma>0$, the following initial condition is stationary, 
$r_{\rm stat} (x)=r(x,t)=r_0(x)$ for all $t$ 
\be \label{stat} 
\!\!  r_{\rm stat} (x) = \frac{1}{2} \tanh ( \frac{\bar \gamma (x-x_s)}{2 T} ) ,
\rho_{\rm stat} (x) = \frac{\bar \gamma}{4 T \cosh^2(\frac{\bar \gamma (x-x_s)}{2 T}  ) }
\ee
and describes a shock at position $x_s$ containing all the particles, which, in this scaling, has
a finite width $O(T/\bar \gamma)$. It perfectly agrees with the large $N$ limit of the 
finite $N$ density profile associated to the IC  
$P_N(\vec x,0) \propto \Psi_0(\vec x)^2 \delta(\bar x-x_s)$,
which in Fourier reads 
$\hat \rho^s_N(K) = \prod_{a=1}^{N-1} (1+ (\frac{K}{N \bar c a})^2)^{-1} \to_{N \to \infty} \frac{\pi K}{\bar \gamma \sinh(\pi K/\bar \gamma)}$ as first calculated in \cite{Rybicki}. Note that the diffusion of the shock center, ${\rm Var} \, x_s(t) \sim
\frac{2 T}{N} t$ is subdominant, and only observable on larger time scales $t \sim N$.

Using the solution to Burgers equation \eqref{solu} one can investigate again the question
of the decay rate towards \eqref{stat}. One finds that fast decaying IC have decay rate $\bar \gamma^2/4$ 
consistent with the large $N$ limit of \eqref{relax1}. On the other hand, one can solve e.g. the case of two packets, $r_0(x) =  \frac{p_1}{2}  \tanh ( \frac{p_1 \bar \gamma}{2} (x-x_1) ) + \frac{p_2}{2} \tanh ( \frac{p_2 \bar \gamma}{2} (x-x_2) )$, with $p_1+p_2=1$, which leads a decay rate  
$p_1 p_2 \bar \gamma^2 \leq \frac{1}{4} \bar \gamma^2$ \cite{SM}. 

In the repulsive case $\gamma=-\bar \gamma>0$, let us investigate the quadratic potential 
$V(x)= \mu \frac{x^2}{2}$. Eq. \eqref{eqrho22} can be put in the form of a Burgers equation
with friction $\mu$ \cite{SM}. We only study the stationary solution. Its support is
the whole real axis, and it takes the scaling form
\be
r_{\rm stat} (x)= \frac{\sqrt{\mu T}}{2 \gamma} \hat r_g(x \sqrt{\frac{\mu}{T}})  ~,~ 
\rho_{\rm stat} (x)= \frac{\mu}{2 \gamma} \hat \rho_g(x \sqrt{\frac{\mu}{T}})  
\ee 
where $\hat r_g(y) = y - R_g(y)$, $\hat \rho_g(y)= 1 - R_g'(y)$ are
a one parameter family of scaling functions indexed by 
$g=\gamma/\sqrt{\mu T}$, which measures the relative strength 
of interaction and potential energies. The odd dimensionless function $R_g(y)$
is the solution of $0 = R_g'' + R_g R_g' -  R_g$ such that
$R_g(y) \simeq y \mp g$ for $y \to \pm \infty$. This autonomous equation
is solved by writing $R'=w(R)$. The solution for $\hat \rho_g(y)$ is 
obtained by eliminating $R$ and $b>0$ between
\bea
&& \!\! \hat \rho_g(y) = - W(- e^{-b-1 - \frac{R^2}{2}}) , 
\int_{0}^R  \frac{du}{1 + W( - e^{-b-1 - \frac{u^2}{2}})}= y \nonumber \\
&& \int_{0}^{+\infty} du \big( \frac{1}{1 + W( - e^{-b-1 - \frac{u^2}{2}})} - 1 \big) = g
\eea 
where $W(z)$ is the main branch of the Lambert function, solution of $z=W e^W$, with $z> e^{-1}$ and $W(e^{-1})=-1$. 
In the limit $g \to 0$, $b$ is large and using $W(z)=\sum_{n \geq 1} \frac{(-n)^{n-1}}{n!} z^n$
one obtains the expansion $\hat \rho_g(y)=\sqrt{\frac{2}{\pi }} g e^{-\frac{y^2}{2}} + O(g^2)$ 
around the Gaussian shape in the absence of interactions $\gamma=0$, see \cite{SM}.
Similarly, for $g \to +\infty$, $b \to 0^-$ and in the scale $y \sim g$ the scaling function
takes a square form
$\hat \rho_g(y) \simeq \theta(1 - \frac{y}{g})$.

{\it Coulomb gas}. The Dean-Kawasaki equation \eqref{eqrho1} 
satisfies detailed balance \cite{Dean,DeanPrivate}, with the energy 
\bea \label{Hrho} 
H[\rho] &=&  \frac{\bar c N^2}{2} \int  dx dx' |x-x'|  \rho(x)  \rho(x') 
+ N \int dx V(x)  \rho(x)  \nonumber \\
&+& T N  \int dx  \rho(x) \log \rho(x) 
\eea 
the last term being the entropy term,
see \cite{SM} for details, and one expects that the system reaches equilibrium with Gibbs measure
${\cal P}_{\rm stat}[\rho] \propto \exp\left(  - H[\rho]/T \right)$. In the first large $N$ limit,
at fixed $c=-\bar c$, with $V(x)=N \tilde V(x)$, one has 
$H[\rho] \simeq N^2 {\cal E}[\rho]$ and only the first two terms in \eqref{Hrho} contribute. 
Consider repulsive interactions $c>0$. 
We have checked \cite{SM} that for convex potentials, minimization of ${\cal E}[\rho]$ gives the
same solution as the asymptotic state \eqref{aga0} of the dynamics \eqref{1storder}. The Coulomb gas in that regime
(in a quadratic well) was studied in the context of Jellium: it was shown that at very low temperature
$T \sim 1/N$, the system almost cristallizes, and that the rightmost particle has 
position fluctuations $O(1/N)$, obtained in \cite{SatyaJellium1,SatyaJellium2}. 

For double well potentials and $c>0$, the density has two supports and 
we find \cite{SM} that the minimizer is only one special member of the family of steady states reached by
the dynamics on the fast time scales $\tau= N t = O(1)$. Not so surprisingly,
barrier crossing is needed to equilibrate the two wells, a process which occurs on much larger
time scales, and would be interesting to study. For $\bar c>0$ the minimizer is a single delta function
packet (shock) at the position of the minimum of $\tilde V(x)$. 

Finally, in the large $N$ limit with fixed $\gamma= N c$ all three term in \eqref{Hrho} are $O(N)$ and
contribute, with $H[\rho] \simeq N {\cal E}_2[\rho]$. Minimization of ${\cal E}_2[\rho]$ recovers the
stationary equation associated to \eqref{eqrho22}.

In conclusion we studied the dynamics of interacting ranked diffusion in 1D using the tools
of the integrable LL model (for attractive interactions) and of the Burgers equation (for both cases,
most useful in the large $N$ limit). Stationary solutions, decay rates and stationary correlations 
were obtained. For IC with fast spatial decay, the decay rate in time is universal and given 
by the LL Hamiltonian spectrum, while it is continuously varying and slower for IC which
decay slowly in space. The Coulomb gas allows to obtain the equilibrium state on
much larger time scales. Since the Burgers equation describes the large $N$ dynamics
(with $\gamma=N \bar c$ fixed) on the same time scales as
the LL model (in imaginary time) it would be interesting to explore 
the correspondence further.

{\it Acknowledgments:} 
I am grateful to S. N. Majumdar for useful interactions at the early stages of this work. 
I thank J. Quastel and L.C. Tsai for discussions on relations to the Burgers equation,
and D. Dean for sharing notes. This research was supported 
by ANR grant ANR-17-CE30-0027-01 RaMaTraF.

%
%

{}

\newpage

.

\newpage

\begin{widetext} 



\setcounter{secnumdepth}{2}

\begin{large}
\begin{center}

Supplementary Material for\\  {\it Ranked diffusion, delta Bose gas and Burgers equation}

\end{center}
\end{large}

\bigskip

We give the details of the calculations described in the main text of the Letter, and display
some of the more lengthy results (in particular the return probabilities and stationary correlations
within the Bethe ansatz, as well as some explicit solutions to Burgers equation). 

\bigskip

\tableofcontents

\section{Associated quantum model}

Inserting $P= \Psi_0 \Psi$ in the Fokker-Planck equation \eqref{FP0} in the text, one finds that $\Psi$ satisfies the imaginary time Schr\"odinger equation
\be
\partial_t \Psi = - {\cal H}'_s  \Psi  = \left( \sum_i \partial_{x_i}^2  + \frac{1}{2} 
 \sum_i \partial_{x_i} F_i - \frac{1}{4} \sum_i F_i^2 \right) \Psi \quad , \quad F_i = V'(x_i )+\bar c \sum_j {\rm sgn}(x_i-x_j)
\ee
For $V(x)=0$ one obtains the Lieb-Liniger Hamiltonian
\bea
{\cal H}'_s =  - \sum_i \partial_{x_i}^2  - 2 \bar c \sum_{1 \leq i < j \leq N} \delta(x_i-x_j) - E_0   \quad , \quad 
E_0= - \frac{\bar c^2}{4} \sum_{i=1}^N (N+1-2 i)^2 = - \frac{\bar c^2}{12} (N^3-N)
\eea 
The ground state of ${\cal H}'_s$ is $\Psi_0$ with zero energy. The standard LL model
has Hamiltonian ${\cal H}_s={\cal H}'_s + E_0$ and has ground state energy $E_0$.

For interacting ranked diffusion (RD) in presence of an external potential $V(x)$ the associated quantum model is not the Lieb-Liniger model anymore, but there are additional one and two body terms
i.e. ${\cal H}_s  \to {\cal H}_s + \delta {\cal H}_s$ with
\bea
&& \delta {\cal H}_s = \frac{1}{4} \sum_i (V'(x_i)^2 - 2 V''(x_i)) + \frac{\bar c}{4} \sum_{i,j} (V'(x_i)-V'(x_j)) {\rm sgn}(x_i-x_j) 
\eea

For RD in the quadratic potential $V(x)=\frac{1}{2} \mu x^2$ the additional terms in the quantum model are
\be
\delta {\cal H}_s = - \frac{\mu N}{2} + \frac{\mu^2}{4} \sum_i x_i^2 + \frac{\mu \bar c}{4} 
\sum_{i,j} |x_i - x_j|
\ee
i.e. a quadratic well and a linear attraction.

For RD in the linear trap $V(x)= \mu |x|$
one finds
\be
\delta {\cal H}_s  = \frac{N}{4} \mu^2 - \mu \sum_i \delta(x_i) + \frac{\bar c \mu}{4} (N^2- (\sum_i {\rm sgn}(x_i))^2)
\ee

\section{Bethe ansatz solution for the ranked diffusion dynamics}

\subsection{General symmetric initial condition} 

We use the quantum mechanical notations for states $|\Psi \rangle$, and their associated
wavefunctions in coordinate basis $\Psi(\vec x)=\langle \vec x | \Psi \rangle$. The
scalar product is $\langle \Phi |\Psi \rangle= \int dx_1 \dots dx_N \Phi^*(\vec x) \Psi(\vec x)$
and the norm square is $||\Psi||^2= \langle \Psi |\Psi \rangle$.
\\

The un-normalized symmetric eigenfunctions of the LL Hamiltonian ${\cal H}_s$ 
in \eqref{LL} with attractive interactions ($\bar c>0$) on the infinite line are built 
\cite{m-65,GaudinBook} by partitioning the $N$ particles 
into a set of $1 \leq n_s \leq N$ bound states called {\it strings} 
each formed by $m_j \geq 1$ particles with $N=\sum_{j=1}^{n_s} m_j$. 
In the sector $x_1 \leq x_2 \leq \dots \leq x_N$ they read
\be \label{def1}
\Psi_\lambda(\vec x) =  \sum_{P \in S_N} A_P \prod_{j=1}^N e^{i \sum_{\alpha=1}^N \lambda_{P_\alpha} x_\alpha} \, , \quad 
A_P=\prod_{1 \leq \alpha < \beta \leq N} \Big(1 + \frac{i  \bar c
}{\lambda_{P_\beta} - \lambda_{P_\alpha}}\Big)\,.
\ee
which involves a sum over permutations $P$ in $S_N$. The rapidities are 
\be\label{stringsol}
\lambda_{j, a}=k_j - \frac{i \bar c}2(m_j+1-2a) \quad , \quad j=1,\dots n_s \quad , \quad a = 1,...,m_j
\ee 
where $k_j$ is a real momentum, the total momentum of the string being $K_j = m_j k_j$. One denotes equivalently $|\Psi_\lambda \rangle \equiv | \Psi_{{\bf k}, {\bf m} } \rangle$ these strings states, 
and 
$\Psi_\lambda(\vec x) =\Psi_{{\bf k}, {\bf m} } (\vec x)$ the corresponding
eigenfunctions, labelled by the set of $k_j,m_j$, $j=1,..n_s$. The eigenenergies are
\be \label{en} 
E_\lambda= \sum_{\alpha=1}^N \lambda_\alpha^2  = E( {\bf k}, {\bf m})  \quad , \quad 
E( {\bf k}, {\bf m})   := \sum_{j=1}^{n_s} m_j k_j^2-\frac{\bar c^2}{12} (m_j^3 - m_j) 
\ee 
and the inverse norms are \cite{CalabreseCauxBosonsPRL,CalabreseCauxBosonsLong,Kirillov}
\bea
&&  \frac{1}{||\Psi_{{\bf k}, {\bf m} }  ||^2} = \frac{\bar c^N}{N!  (\bar c L)^{n_s} } \Phi({\bf k}, {\bf m}) \prod_{j=1}^{n_s} \frac1{m_j^{2}}  
\quad , \quad  \Phi({\bf k}, {\bf m}) = \prod_{1\leq i<j\leq n_s} \frac{4(k_i-k_j)^2 +(m_i-m_j)^2 \bar c^2}{4(k_i-k_j)^2 +(m_i+m_j)^2 \bar c^2}  \label{norm}
\eea
This formula is valid on a ring of size $L$ as $L \to +\infty$. When summing over states
the factors of $L$ cancel in that limit, upon 
using the quantification of the total string momenta $\sum_{k_j} \to L m_j  \int \frac{dk}{2 \pi}$.
See e.g. \cite{we-flatlong} for more details (same conventions). 
\\

On the infinite line this leads to the following exact expression for the ranked diffusion probability, which we denote
$P_N(\vec x,t)$, with a symmetric initial condition $P_N(\vec x,0)$
\be \label{PBethe} 
 P_N(\vec x,t)  = \Psi_0(\vec x)  {\cal Z}_N(\vec x,t) e^{E_0 t}
\ee
where the sum over states in \eqref{Pt} in the text takes the form 
\bea \label{sum2} 
 {\cal Z}_N(\vec x,t)  =
\sum_{n_s=1}^N \frac{\bar c^{N-n_s}}{n_s! N!}  \prod_{j=1}^{n_s} \sum_{m_j \geq 1} \int \frac{dk_j}{2 \pi m_j}  
\delta_{N, \sum_{j=1}^{n_s} m_j}
\Phi({\bf k}, {\bf m}) e^{- t E( {\bf k} , {\bf m})} 
\Psi_{{\bf k}, {\bf m} } (\vec x) 
 \langle \Psi(0) | \Psi_{{\bf k}, {\bf m} }  \rangle
\eea 
Here the state $|\Psi(0) \rangle$ is the initial condition of the quantum evolution and reads
in the coordinate basis, together with the overlap 
\be  \label{overlap00} 
\langle \vec x| \Psi(0) \rangle=\frac{P_N(\vec x,0)}{\Psi_0(\vec x)} \quad , \quad 
 \langle \Psi(0) | \Psi_{{\bf k}, {\bf m} }  \rangle = \int d^N \vec x 
 \, \frac{P_N(\vec x,0)}{\Psi_0(\vec x)} \Psi_{{\bf k}, {\bf m} }(\vec x) 
\ee
We will assume here that these overlap integrals exist, which assumes that the spatial decay
of $P_N(\vec x,0)$ at large $\vec x$ is sufficiently fast, as discussed in the text.
\\

{\it Ground state}. The ground state is a single $N$-string with $k_1=0$ and $m_1=N$, i.e 
\be 
\Psi_0(\vec x)  = \Psi_{0,N}(\vec x)  = N! e^{\frac{\bar c}{2} \sum_{a=1}^N (N+1-2 a) x_{(a)} }
= N! e^{- \frac{\bar c}{2} \sum_{1 \leq i<j \leq N} |x_i-x_j|} 
\ee 
and energy $E_0=E(0,N)= -\frac{\bar c^2}{12} (N^3 - N)$. Its square norm is
$||\Psi_0 ||^2 = N! N^2 L \bar c^{1-N}$, where the factor of $L$ comes from integration over the (free) center
of mass coordinate. 

\subsubsection{Ground state manifold}

There is a manifold of eigenstates which plays a special role, let us call it the ground state manifold.
It corresponds to $n_s=1$ and 
consists of superpositions of single $N$-string states but with arbitrary momentum, with associated wave functions
$\Psi_{k,N}(\vec x)=\Psi_0(\vec x) e^{i k \sum_i x_i}$. 
It has two properties. First if the initial condition belongs to this manifold it remains inside and its 
dynamics then describes the time evolution of the center of mass. Second, for any initial condition, 
we expect that the evolution will converge at large time to this manifold.

Separating the term $n_s=1$ in \eqref{sum2} and inserting
into \eqref{PBethe} we see that we can write
\be  \label{sumP} 
P(\vec x,t)  = P^{n_s=1}(\vec x,t)  + P^{n_s \geq 2}(\vec x,t) 
\ee 
where the first term evolves inside the ground state manifold and reads 
\bea \label{sum3} 
 P^{n_s=1}(\vec x,t)  =  \Psi_0(\vec x)^2
\frac{\bar c^{N-1}}{N!}  \int \frac{dk}{2 \pi N}  
 e^{- t N k^2}  e^{i k \sum_i x_i}  \langle \Psi(0) | \Psi_{k,N} \rangle
\eea 
and where the overlap is
\be \label{overlapground} 
 \langle \Psi(0) | \Psi_{k,N} \rangle = \int d^N\vec x P(\vec x,0) e^{- i k \sum_i x_i} 
\ee 

Suppose first that the initial condition is of the product form
\be  \label{initdec} 
P_N(\vec x,0)= \frac{\bar c^{N-1}}{N! N^2}  \Psi_0(\vec x)^2 \, p_0(\frac{1}{N} \sum_i x_i) 
\ee 
i.e. it is stationary in the particle relative positions, with a decoupled form for the center of mass.
Then it is easy to see that it remains of this form.
Indeed, defining the Fourier transform $\tilde p_0(K)$ of the initial PDF $p_0(x)$ for the center of mass as
\be 
p_0(x) = \int \frac{dK}{2 \pi} e^{i K x}  \tilde p_0(K)  
\ee 
one can rewrite the initial condition as (setting $K= N k$ the total string momentum)
\be \label{dec2} 
P_N(\vec x,0)= \frac{\bar c^{N-1}}{N! N^2}  \Psi_0(\vec x)^2 \int \frac{dk}{2 \pi} e^{i k \sum_i x_i} \hat p_0(k) \quad , \quad 
\hat p_0(k) = N \tilde p_0(N k)
\ee 
and one sees that $P_N(\vec x,0)/\Psi_0(x)$ belongs to the ground state manifold. Let us check that its evolution is indeed simple, from the above formula \eqref{sum3}. Let us compute the overlap \eqref{overlapground} from \eqref{dec2}. One obtains 
\be 
\langle \Psi(0) | \Psi_{k,N} \rangle = \frac{\bar c^{N-1}}{N! N^2}  \int d^N\vec x \int \frac{dq}{2 \pi} 
\Psi_0(\vec x)^2  e^{i (q-k) \sum_i x_i} \hat p_0(q) = \frac{1}{N} \hat p_0(k) 
\ee 
where we have used that $\delta_{k,0}=\frac{2 \pi}{L N} \delta(k)$, hence that
\be 
\int d^N \vec x \Psi_0(\vec x)^2 e^{i k \sum_i x_i} = ||\Psi_0||^2 \delta_{k,0} = 
||\Psi_0||^2 \frac{2 \pi}{L N} \delta(k) 
= N! N \bar c^{1-N} 2 \pi \delta(k) 
\ee 
Inserting the overlap in \eqref{sum3} leads to 
\be 
P_N(\vec x,t)= \frac{\bar c^{N-1}}{N! N^2}  \Psi_0(x)^2
  \int \frac{dk}{2 \pi} e^{i k \sum_i x_i} \hat p_0(k) e^{- N k^2 t}
= \frac{\bar c^{N-1}}{N! N^2}  \Psi_0(\vec x)^2 \times p_t(\frac{1}{N} \sum_i x_i) 
\ee 
where $p_t(x)$ is the PDF of the center of mass at time $t$
\be 
p_t(x) = \int \frac{dK}{2 \pi} e^{i K x}  \tilde p_0(K) e^{- \frac{K^2}{N} t} 
\ee 
The center of mass performs diffusion with variance $\frac{2}{N} t$. 
Note that if $\int dx \, p_0(x)=\tilde p_0(0)=1$ then $P_N(\vec x,t)$ is normalized to unity for all $t$
(see subsection below).
\\

Consider now an initial condition which is not of this form. As mentionned in the text there is a gap in the energy spectrum between states with $n_s \geq 2$ and with
$n_s=1$, of value $\gamma_N = \frac{\bar c^2}{4} N(N-1)$. Hence we expect a
convergence of the form $\sim e^{-\gamma_N t}$ towards the ground state manifold,
i.e. $P_N(\vec x,t) \simeq P^{n_s =1}(\vec x,t)$ for times $t \gg 1/\gamma_N$. 
As discussed in the text this holds provided the overlaps are convergent integrals,
i.e. if the spatial decay of $P_N(\vec x,0)$ at infinity is fast enough.

As an example let us consider as in the text the "droplet" initial condition
\be \label{drop1} 
P_N(\vec x,0)= \prod_{i=1}^N \delta(x_i) 
\ee 
which is normalized to unity.
The overlap is now  $\langle \Psi(0) | \Psi_{k,N} \rangle =1$ and 
one obtains from the above the large time behavior
\be 
P_N(\vec x,t) \simeq P^{n_s =1}(\vec x,t) = 
\frac{\bar c^{N-1}}{N! N^2}  \Psi_0(x)^2
\, p_t(\frac{1}{N} \sum_i x_i) \quad , \quad 
p_t(x)= \frac{\sqrt{N}}{\sqrt{4 \pi t} } e^{- N \frac{x^2}{4 t} } 
\ee 
Note that at intermediate time all excited states contribute and the dynamics is more complicated.

\subsubsection{Separation of the center of mass motion}

It is possible to separate completely the center of mass coordinate $x$ from the relative coordinate variables. If the following factorized form holds at time $t=0$, it holds at all time 
\be \label{facto} 
P_N(\vec x,t) = p_t(x) P_t(\vec y) \quad , \quad x= \frac{1}{N} \sum_{i=1}^N x_i \quad , \quad y_i=x_i-x
\ee 
The normalization condition $\int d^N \vec x P_N(\vec x,t)  = 1$ is satisfied if
\be 
\int dx \, p_t(x) =1 \quad , \quad \int d^N \vec y \, \, \delta(\frac{1}{N} \sum_{i=1}^N y_i) P_t(\vec y) =1
\ee 
As can be seen by using that $\int d^N \vec x  = \int dx \int d^N \vec x \delta(x- \frac{1}{N} \sum_{i=1}^N x_i)
= \int dx \int d^N \vec y \delta( \frac{1}{N} \sum_{i=1}^N y_i)$ upon shifting $x_i = y_i + x$. 
\\

To show \eqref{facto} for all time, note that setting $x_i = x + y_i$ in the expression for the eigenstates \eqref{def1}, \eqref{stringsol} they take the form
\be 
\Psi_{{\bf k}, {\bf m}}(\vec x)= e^{i \sum_{j=1}^{n_s} k_j m_j x}  \Psi_{{\bf k}, {\bf m}}(\vec y)
\ee 
Inserting $\int dx \delta(x- \frac{1}{N} \sum_i x_i)=1$ in the expression of the overlap \eqref{overlap00} one obtains 
\bea  
&&  \langle \Psi(0) | \Psi_{{\bf k}, {\bf m} }  \rangle =
  \tilde p_0(\sum_{j=1}^{n_s} m_j k_j)   \langle \Psi(0) | \Psi_{{\bf k}, {\bf m} }  \rangle_y \\
&& \tilde p_0(K):=\int dx e^{i K x} p_0(x)  \quad , \quad
 \langle \Psi(0) | \Psi_{{\bf k}, {\bf m} }  \rangle_y := \int d^N \vec y \, \delta(\frac{1}{N} \sum_{i=1}^N y_i)  \, \frac{P_0(\vec y)}{\Psi_0(\vec y)}   \Psi_{{\bf k}, {\bf m} }(\vec y)
 \eea
Inserting into \eqref{sum2} one obtains
\bea \label{sum4} 
&&  P_N(\vec x,t)  =  \int \frac{dK}{2 \pi} e^{i K x}   \tilde p_0(K) \\
 && 
\times  \Psi_0(\vec y)
\sum_{n_s=1}^N \frac{\bar c^{N-n_s}}{n_s! N!}  \prod_{j=1}^{n_s} \sum_{m_j \geq 1} \int \frac{dk_j}{2 \pi m_j}  
\delta_{N, \sum_{j=1}^{n_s} m_j} (2 \pi) \delta(K- \sum_{j=1}^{n_s} m_j k_j) 
\Phi({\bf k}, {\bf m}) e^{- t (E( {\bf k} , {\bf m})-E_0)} 
\Psi_{{\bf k}, {\bf m} } (\vec y) 
 \langle \Psi(0) | \Psi_{{\bf k}, {\bf m} }  \rangle_y \nonumber 
\eea 
Now one notes that shifting the integration variables as $k_j = k + q_j$, all terms inside the integral
are invariant except the constraint $\delta(K- \sum_{j=1}^{n_s} k_j m_j) \to \delta(K- N k - \sum_{j=1}^{n_s} m_j q_j )$ 
and the energy $E( {\bf k} , {\bf m}) = N k^2 + 2 k \sum_{j=1}^{n_s} m_j q_j + \sum_{j=1}^{n_s}  m_j q_j^2$. 
If we choose $k=K/N$ the expression for the energy decouples and one finally obtains \eqref{facto} with
\bea \label{sum5} 
&&  p_t(x) =  \int \frac{dK}{2 \pi} e^{i K x}   \tilde p_0(K) e^{- \frac{K^2}{N} t} \\
&& P_t(\vec y) = 
\Psi_0(\vec y)
\sum_{n_s=1}^N \frac{\bar c^{N-n_s}}{n_s! N!}  \prod_{j=1}^{n_s} \sum_{m_j \geq 1} \int \frac{dq_j}{2 \pi m_j}  
\delta_{N, \sum_{j=1}^{n_s} m_j} (2 \pi) \delta(\sum_{j=1}^{n_s} m_j q_j) 
\Phi({\bf q}, {\bf m}) e^{- t (E( {\bf q} , {\bf m})-E_0)} 
\Psi_{{\bf q}, {\bf m} } (\vec y) 
 \langle \Psi(0) | \Psi_{{\bf q}, {\bf m} }  \rangle_y \nonumber
 \eea

\subsection{Probability of presence at the origin} 

The formula \eqref{PBethe}, \eqref{sum2} are exact but the Bethe eigenfunctions become quite complicated for higher values of $n_s$. A simpler observable is the probability of presence at 
the origin. \\

{\it Droplet initial condition}. Let us consider again the initial condition \eqref{drop1} where all particles
are at $x=0$ at time $t=0$ and ask the probability at time $t$, 
$P(\vec x=0, t) (dx)^N$ that they are all in a small volume $(dx)^N$ around the origin.
In that case it is a {\it return probability}. 

We insert in \eqref{PBethe}, \eqref{sum2} the exact value of the Bethe wavefunctions
\eqref{def1} at coinciding points $\Psi_\lambda(\vec 0) = N!$ and of the 
overlap with the initial state, which consequently equals unity, $\langle \Psi(0) | \Psi_{{\bf k}, {\bf m} }  \rangle =1$, and obtain
\bea \label{zbethe2} 
&& P(\vec x=0,t)  = N! e^{E_0 t} \sum_{n_s=1}^N \frac{\bar c^{N-n_s}}{n_s!} 
 \prod_{j=1}^{n_s} \sum_{m_j \geq 1} \int \frac{dk_j}{2 \pi m_j}   \delta_{N, \sum_{j=1}^{n_s} m_j} e^{t \sum_{j=1}^{n_s} [ \frac{\bar c^2}{12} (m_j^3-m_j) - m_j k_j^2]}
\prod_{1 \leq i < j \leq n_s} \frac{4(k_i-k_j)^2 + (m_i-m_j)^2 \bar c^2}{4(k_i-k_j)^2 + (m_i+m_j)^2 \bar c^2} \nn
\eea 

For instance one finds $P_1(0,t) = \int \frac{dk}{2 \pi} e^{- k^2 t } = \frac{1}{\sqrt{4 \pi t} }$ and
\be
 P_2(\vec 0,t) = \bar c \int \frac{dk}{2 \pi} e^{-2 k^2 t} + e^{- \frac{1}{2} \bar c^2 t}  
\int \frac{dk_1}{2 \pi} \int \frac{dk_2}{2 \pi}  e^{- k_1^2 t - k_2^2 t} 
\frac{(k_1-k_2)^2}{(k_1-k_2)^2 + \bar c^2} 
 = \frac{2 e^{-\frac{\bar c^2 t}{2}}+\sqrt{2 \pi } \bar c \sqrt{t}
   \left(\text{erf}\left(\frac{\bar c
   \sqrt{t}}{\sqrt{2}}\right)+1\right)}{8 \pi  t}
\ee
where ${\rm erf}(z)=\frac{2}{\sqrt{\pi}} \int_0^z dt e^{-t^2}$, 
and so on for higher $N$. This coincides with the result for the moments of the KPZ equation
with droplet initial conditions, $P_N(\vec 0,t)= e^{E_0 t} \overline{Z(0,t)^N}= e^{E_0 t} \overline{e^{N h(0,t)}}$, as obtained in \cite{PLDdroplet} (see Eqs. (9)-(11) there). Here $h(x,t)$ is the height 
field which obeys the KPZ equation $ \partial_t h = \partial_x^2 h + (\partial_x h)^2 + \sqrt{2 \bar c} \eta(x,t)$  
where $\eta(x,t)$ a standard white noise (which is averaged over). Denoting
$Z(x,t)=e^{h(x,t)}$ the droplet initial condition is $Z(x,t=0)=\delta(x)$. 
\\

For general $N$ the return probability 
$P_N(\vec 0,t)$ interpolates between $P_N(\vec 0,t) \simeq P_1(0,t)^N$ at short time and 
$P_N(\vec 0,t) \simeq \bar c^{N-1} N! \frac{1}{N \sqrt{4 \pi N t}}$ 
at large time. 
\\

{\it Flat initial condition}. Another initial condition of interest is
\be 
P_N(\vec x,t=0)= C_N \Psi_0(\vec x) 
\ee 
If we choose $C_N = (\bar c/2)^{N-1}/(N! N^2 L)$, then it is normalized to $\int d^N \vec x P(\vec x,0)=1$,
i.e. with a uniform density for the center of mass. In the KPZ context corresponds to the flat IC
$Z(x,t=0)=1$ with ${\cal Z}_N(\vec 0,t)= C_N \overline{Z(0,t)^N}$, hence the probability density of
presence at the origin 
\be 
P_N(\vec 0,t) = N! e^{E_0 t} C_N \overline{Z(0,t)^N}
\ee 
The moments $\overline{Z(0,t)^N}$ have been obtained in \cite{we-flat,we-flatlong}.
For $N \leq 4$ explicit formula where given in \cite{we-flatlong,flat-shorttime}.
Using these formula we obtain
\bea 
P_2(\vec 0,t) = 2 C_2  \left( 1+ {\rm erf}(\bar c \frac{\sqrt{t}}{\sqrt{2}} ) \right)
\eea 
which increases by a factor of $2$ between $t=0$ and $t=+\infty$. 
Next we obtain
\be \label{eqz3}
 P_3(\vec 0,t) = 6 C_3 \left(  
4 - 2 e^{- \frac{3}{2} \bar c^2 t }  -  2  {\rm erfc}(\bar c \sqrt{2 t}) 
+e^{- \frac{3}{2} \bar c^2 t } {\rm erfc}(\bar c \frac{\sqrt{t}}{\sqrt{2}})
 \right)
\ee
which increases by a factor of $4$ between $t=0$ and $t=+\infty$.
Finally one has 
\bea
&& P_4(\vec 0,t) = 4! C_4 \bigg(  8 - 8 e^{- 3 \bar c^2 t} - 4   {\rm erfc}(\frac{3}{ \sqrt{2}} \bar c \sqrt{t}) + 8 e^{-3 \bar c^2 t}  {\rm erfc}(\bar c \sqrt{t}) - 4  {\rm erfc}(2 \bar c \sqrt{t})
  \nonumber\\
&& +  48 e^{- 5 \bar c^2 t}  \int_0^\infty dx 
\frac{\left(\frac{2 x+1}{\sqrt{4 x+1}}-\frac{\sqrt{2 x+1}}{x+1}\right) e^{-2 \bar c^2 t x}}{4 \pi  (4 x (x+3)+5)}  \bigg)
\eea
which increases by a factor of $8$ between $t=0$ and $t=+\infty$.
For general $N$ similar formula can be obtained and the increase between 
$t=0$ and $t=+\infty$ is $2^{N-1}$. 

Note that the flat IC is the limiting case where the overlap integrals exist.
\\

Finally, there are also results for Brownian initial conditions for the KPZ equation,
These corresponds to more exotic initial conditions for the RD problem
\be
P_N(\vec x,0) \propto \Psi_0(\vec x) 
\mathbb{E} [ e^{\sum_{j=1}^N \sqrt{\bar c} B(x_j)}  ] 
\ee 
where $B(x)$ is a two sided Brownian motion in $x$ with $B(0)=0$, 
with possibly two different drifts $w_{L,R}$ on each side. It can be
made more explicit using 
and 
\bea
&& \mathbb{E}  e^{\sum_{j=1}^p B(-y_j)}  = e^{- \frac{\bar c}{2} \sum_{j=1}^p (2 j - 1) y_j} \quad , \quad y_1<..< y_p<0 \\
&& \mathbb{E}  e^{\sum_{j=1}^p B(y_j)}  = e^{\frac{\bar c}{2} \sum_{j=1}^p (2 p - 2 j + 1) y_j} \quad , \quad 0<y_1<..< y_p
\eea
We have not attempted to study this case, but it may be of interest. The overlap integrals will require analytic
continuations, since the spatial decay is slower than for the flat IC. 

\subsection{Correlation functions}

Consider an observable, for instance the density (here normalized to $N$ so we use the notation $\tilde \rho$ 
of the main text)
\be  \label{tilde} 
{\tilde \rho}(x;\vec x) = \sum_i \delta(x-x_i) 
\ee 
Its expectation value in the stationary state of the RD system, i.e. 
$P_N(\vec x,0)=\Psi_0(\vec x)^2/||\Psi_0||^2$ (which is normalized to unity), 
is given by
\be 
\langle {\tilde \rho}(x;\vec x(0)) \rangle = \int d^N \vec x {\tilde \rho}(x;\vec x) \frac{\Psi_0(\vec x)^2}{||\Psi_0||^2}
=  \frac{\langle \Psi_0 | \hat \rho(x,0) | \Psi_0 \rangle}{||\Psi_0||^2} 
\ee 
where $\hat \rho(x,0)$ is the density operator corresponding to the observable \eqref{tilde}. 
Hence it is equal to the quantum expectation of the density operator in the
ground state of the LL model. Note that in that state the center of mass has uniform probability 
over the system hence the result is simply $N/L$. We will study below the more
interesting case where the center of mass position is fixed. 

Consider now a time dependent correlation in the RD system. It can be expressed as a quantum matrix element in imaginary time. Indeed one has
\bea \label{corrderivation} 
&& \langle {\tilde \rho}(x,t) {\tilde \rho}(y,0) \rangle =
\langle {\tilde \rho}(x;\vec x(t)) {\tilde \rho}(y;\vec x(0)) \rangle 
= \int d^N \vec x \int d^N \vec y {\tilde \rho}(x;\vec x) G_{FP}(\vec x,\vec y, t) {\tilde \rho}(y;\vec y) 
P_N(\vec y,0) \\
&& = e^{E_0 t} \int d^N \vec x \int d^N \vec y \Psi_0(\vec x) {\tilde \rho}(x;\vec x)  G_{s}(\vec x,\vec y, t) 
{\tilde \rho}(y;\vec y) \langle \vec y |\Psi(0) \rangle \\
&& = \langle \Psi_0 | \hat \rho(x,t) \hat \rho(y,0) | \Psi(0) \rangle
\eea 
where $\hat \rho(x,t)$ is the density operator in the Heisenberg representation, i.e.
$\hat \rho(x,t) = e^{ \hat H_s t} \hat \rho(x,0)  e^{ - \hat H_s t}$ in imaginary time
(with $t = i \tau$ and $\tau$ is quantum real time). 
\\

For instance consider the time correlation of the density in the stationary state of the RD system,
which corresponds to the 
initial (unnormalized) quantum state $\langle \vec y | \Psi(0) \rangle = \Psi_0(\vec y)/||\Psi_0||^2$.
This gives 
\bea 
&& 
C(x-y,t) := \langle {\tilde \rho}(x,t) {\tilde \rho}(y,0) \rangle =
 \frac{\langle \Psi_0 | \hat \rho(x,t) \hat \rho(y,0) | \Psi_0 \rangle}{||\Psi_0||^2} 
\eea 
This relation extends to any number of space-times points, although we will focus here on two points.
It extends in fact to any operator, for instance it also applies to correlations of the fields operator
$\Phi^\dagger(x)$ and $\Phi(x)$ which insert an additional particle, or respectively remove it. 
The correlations of this operator were computed in \cite{CalabreseCauxBosonsPRL,CalabreseCauxBosonsLong} 
but we will not study it here. 
\\

The stationary dynamics are thus related in both systems.  To make it explicit one defines
the form factors 
$\Sigma^\rho_\lambda$
of the density operator between the ground state and an arbitrary eigenstate
 $|\Psi_\lambda \rangle$ of the LL Hamiltonian 
\be \label{ffdef} 
 \langle \Psi_0 |\hat \rho(x)|\Psi_\lambda \rangle = \Sigma^\rho_\lambda e^{i K_\lambda x}
\ee
where $K_\lambda$ is the total momentum of the eigenstate $|\Psi_\lambda \rangle$.
The two point stationary density correlation function $C(x,t)$ in the 
RD system can then be expressed using these form factors as a sum over intermediate eigenstates $\Psi_\lambda$
\bea \label{sum10} 
&& C(x,t)=  \frac{\langle \Psi_0 | \hat \rho(x,t) \hat \rho(0,0) | \Psi_0 \rangle}{||\Psi_0||^2}  
 =  \sum_\lambda \frac{ |\Sigma^\rho_\lambda|^2 }{||\Psi_0||^2 ||\Psi_\lambda||^2} 
 e^{- (E_\lambda - E_0) t + i K_\lambda x } \\
 && = 
 \sum_{n_s=1}^N \frac{\bar c^{2N-1-n_s}}{n_s! N!^2 N^2 L}  \prod_{j=1}^{n_s} \sum_{m_j \geq 1} \int \frac{dk_j}{2 \pi m_j}  
\delta_{N, \sum_{j=1}^{n_s} m_j}
\Phi({\bf k}, {\bf m})  |\Sigma^\rho_{{\bf k}, {\bf m} } |^2 e^{i x \sum_{j=1}^{n_s} m_j k_j - t (E( {\bf k} , {\bf m})-E_0)} 
\label{sum100} 
\eea 
The explicit expressions for these form factors were obtained by Calabrese and Caux 
\cite{CalabreseCauxBosonsPRL,CalabreseCauxBosonsLong}
and we can thus apply their results. 
One has, in our conventions, 
\be \label{FFCC} 
|\Sigma^\rho_\lambda| = \frac{N!}{\bar c^N} \frac{(\sum_j m_j k_j)^2}{\bar c} N! (N-1)! \prod_{j=1}^{n_s} H_{m_j}(k_j/\bar c) 
\quad , \quad 
H_m(x)= |\frac{\Gamma( \frac{N-m}{2} + i x)}{\Gamma( \frac{N+m}{2} + i x)}|^2
\ee 

Let us give the explicit result for $N=2$ and perform some checks. We write
\be \label{fourier} 
 \langle {\tilde \rho}(x,t) {\tilde \rho}(y,0) \rangle= \int \frac{dK}{2 \pi} e^{i K (x-y)} C(K,t)
 \ee 
Performing the sum over states in \eqref{sum100} 
we obtain the correlation for $N=2$ in Fourier space 
\bea \label{resN2} 
C(K,t)= \frac{4}{L} \frac{\bar c^4}{(\frac{1}{4} K^2 + \bar c^2)^2} e^{- \frac{1}{2}  t K^2}  
+  \frac{64 \bar c^3 e^{- \frac{1}{2} \bar c^2 t}}{L}  K^4 e^{- \frac{1}{2}  t K^2} 
 \int \frac{dq}{2 \pi} \frac{e^{- \frac{1}{2}  t q^2} }{((q+K)^2 + \bar c^2)^2
((q-K)^2 + \bar c^2)^2 } \frac{q^2}{q^2+\bar c^2} 
\eea 
where the first term arises from the intermediate excited state $\lambda$ 
being a $2$-string  (with total momentum $K=2 k_1$) and the second term
from the two $1$-strings (with total momentum $K=k_1+k_2$).

There are two important sum rules which allow to check the result. 
The first one is obtained by integrating \eqref{fourier} over $x$ 
which gives 
\be 
C(K=0,t)= \int dx \langle {\tilde \rho}(x,t) {\tilde \rho}(0,0) \rangle = N \langle {\tilde \rho}(0,0) \rangle = 2 \frac{2}{||\Psi_0||^2} \int dy_2 \Psi_0(y,y_2)^2 = \frac{4}{L}
\ee 
and is indeed satisfied by \eqref{resN2}. The second is the so-called f-sum rule 
\cite{CalabreseCauxBosonsPRL,CalabreseCauxBosonsLong}. It is obtained by noticing
that $[[ \hat H , {\tilde \rho}_K], {\tilde \rho}_{-K} ]= - \sum_j [[ \partial_{x_j} , e^{i K x_j}], e^{- i K x_j} ]
= - 2 N K^2$ where ${\tilde \rho}_K= \sum_j e^{i K x_j}$. Taking its expectation in the ground state leads to
\be
- L \partial_t C(K,t) |_{t=0} = \sum_\lambda (E_\lambda-E_0) 
\frac{| \langle \Psi_\lambda |\hat \rho_K|\Psi_0 \rangle|^2}{||\Psi_\lambda||^2 ||\Psi_0||^2} 
= N K^2
\ee
From \eqref{resN2} one finds that the sum of the two contributions indeed simplifies into
\be 
- L \partial_t C(K,t) |_{t=0}   = 2 K^2 \frac{\bar c^4}{(\frac{1}{4} K^2 + \bar c^2)^2} 
+  2 K^4 
\frac{K^2+8 \bar c^2}{16 \left(K^2+4 \bar c^2\right)^2} = 2 K^2
\ee

\subsubsection{$N$-string contribution to the density correlation } 

For general $N$ there are many terms in the sum \eqref{sum100}. 
We can see already on \eqref{resN2} that in the large time limit at fixed $K$ 
the $2$-string (the ground state manifold) dominates. 

For the $N$-string for any $N$, of total momentum $N k$, one has from \eqref{FFCC} 
\be \label{SigmaN} 
\Sigma_k^\rho = \frac{N! N^3}{\bar c^{N-1}} \frac{1}{\prod_{a=1}^{N-1} (1 + (\frac{k}{\bar c a} )^2)}
\ee 

This leads to the following contribution of the ground state manifold to the density correlation function 
\be 
C(K,t)|_{\rm N-string} = \frac{ N^2 e^{- \frac{1}{N} K^2 t} }{L \prod_{a=1}^{N-1} (1 + (\frac{K}{N \bar c a} )^2)^2 }
\ee 
which saturates the sum rule $C(0,t)=N^2/L$ and which we expect to dominate in the large time limit
for any $N$ at fixed $K$. In the next subsection we give an interpretation for this result in terms of shocks. 
\\

{\it Large $N$ limit }. In the large $N$ limit with $g=N \bar c$ fixed one has
\be 
C(K,t)|_{\rm N-string}  \simeq \frac{ N^2 (\pi K/g)^2 e^{- \frac{1}{N} K^2 t} }{L \sinh(\pi K/g)^2 }
\ee 
which saturates the sum rule \cite{CalabreseCauxBosonsPRL,CalabreseCauxBosonsLong}
\be 
\int \frac{dK}{2 \pi} C(K,t)|_{\rm N-string}  \simeq \frac{N^3 \bar c}{6 L} 
\ee

In Ref. \cite{CalabreseCauxBosonsPRL,CalabreseCauxBosonsLong} more excited
states are considered, i.e. the two string states $N-M$, $M$. At fixed $N$, this captures some of
the finite time behavior (but not all since there are many more possible excited states). 
However it was found that including already simply $M=1$ saturates the $f$ sum rule at large $N$. 
That may be an indication of how to recover the full Burgers dynamics from the LL Hamiltonian,
but we leave this to future investigations.

\subsubsection{Localized center of mass}

One would also like to study the correlation functions in the RD system with 
a slightly more general initial condition of the type \eqref{initdec} (i.e. in the ground state manifold)
with a non-uniform distribution for the center of mass position
\be \label{init3} 
P_N(\vec x,0) = \frac{\bar c^{N-1}}{N! N^2}  \Psi_0(\vec x)^2 \, p_0(\frac{1}{N} \sum_i x_i) =
 \frac{N L}{ ||\Psi_0||^2} 
\int \frac{dk}{2 \pi}  \tilde p_0(N k)  \Psi_0(\vec x) \Psi_k(\vec x)  
\ee 
where in the last equality we used \eqref{dec2} and $\Psi_k(\vec x) = e^{i k \sum_i x_i} \Psi_0(\vec x)$, as well as the Fourier transform $\tilde p_0(K)=\int dx e^{- i K x} p_0(x)$. 
A case of special interest is $p_0(x)=\delta(x)$ 
where the center of mass is at a fixed position, with $\tilde p_0(K)=1$. 
\\

The mean density is already non trivial, and can be expressed using the form factor $\Sigma^\rho_k$ associated to the
$N$-string with total momentum $K=N k$
\bea 
\langle {\tilde \rho}(x,0) \rangle = \int d^N \vec x \, {\tilde \rho}(x;\vec x) P_N(\vec x,0) =  N L \int \frac{dk}{2 \pi}  \tilde p_0(N k)  
 \frac{\langle \Psi_0 | \hat \rho(x) | \Psi_k \rangle}{||\Psi_0||^2}  = N L \int \frac{dk}{2 \pi}  \tilde p_0(N k)  e^{i N k x} 
 \frac{\Sigma^\rho_k}{||\Psi_0||^2} \\
\eea 
Inserting the result \eqref{SigmaN} we obtain
\be \label{densityLLN} 
\langle \tilde \rho(x,0) \rangle = N \langle \rho(x) \rangle = N \int \frac{dK}{2 \pi} \tilde p_0(K) e^{i K x} 
\frac{1}{\prod_{a=1}^{N-1} (1 + (\frac{K}{N \bar c a} )^2)}
\ee 
which is properly normalized $\int dx \langle {\tilde \rho}(x) \rangle = N$, since $\int dx p(x)=\tilde p_0(0)=1$. 
In the particular case of $p_0(x)=\delta(x)$ (i.e. center of mass fixed at the origin) this is the result
obtained by Rybicki in the context of the self-gravitating gas \cite{Rybicki}, as recalled in the text.
Note that it is simply obtained from the density form factor. 
\\

In the large $N$ limit at fixed $\bar c$ one finds $\langle \rho(x,0) \rangle = p_0(x)$ (i.e the width of the 
packet can be neglected).
\\

In the large $N$ limit at fixed $\bar \gamma = N \bar c$, using 
\be 
\frac{1}{\prod_{a=1}^{N-1} (1 + (\frac{x}{a} )^2)} =  \frac{\pi  x}{\sinh(\pi  x)} +\frac{\pi  x^3}{N \sinh(\pi  x)}  + O(N^{-2}) 
\ee 
One finds 
\be \label{densityLL} 
\langle \rho(x,0) \rangle \simeq  \int \frac{dK}{2 \pi} \tilde p_0(K) e^{i K x} \frac{\pi K}{\bar \gamma \sinh(\frac{\pi}{\bar \gamma}  K)} 
\ee 
In the case of fixed center of mass position, $p(x)=\delta(x)$, inverting the Fourier transform one finds
\be
\langle \rho(x,0) \rangle \simeq \frac{\bar \gamma}{4 \cosh^2( \frac{\bar \gamma}{2} x)} 
\ee
This is precisely the density profile obtained below in \eqref{stat2} as the stationary solution of the Burgers equation 
and corresponding to a single shock centered at $x=0$ and containing all the particles. 

When the center of mass has initial distribution $p_0(x)$ both the finite $N$ result 
\eqref{densityLLN} and its large $N$ limit \eqref{densityLL} can be written as a convolution
\be 
\langle \rho(x) \rangle \simeq \int dx_s \, p_0(x_s) \rho^s_N(x-x_s) 
\quad , \quad \rho^s_{\infty} (x) =  \frac{\bar \gamma}{4 \cosh^2( \frac{\bar \gamma}{2} (x-x_s))} 
\ee 
which can be interpreted as a superposition of a group of $N$ particles centered at random positions $x_s$.
Here $\rho^s_N(x)$ is the finite $N$ density profile of the group, and $\rho^s_\infty(x)$ 
its large $N$ limit, which becomes the stationary shock in the Burgers picture. 
\\

{\it Correlation at different times}. Reproducing the steps in \eqref{corrderivation} we now obtain the two time density correlation of the RD system
with the initial condition \eqref{init3} as
\be 
 \langle \tilde \rho(x,t) \tilde \rho(y,0) \rangle = \frac{N L}{ ||\Psi_0||^2} 
\int \frac{dk}{2 \pi}  \tilde p_0(N k) 
 \langle \Psi_0 | \hat \rho(x,t) \hat \rho(y,0) | \Psi_k \rangle 
\ee 
It can be written again as a sum involving form factors of the LL model
\bea
\langle \tilde \rho(x,t) \tilde \rho(y,0) \rangle = \frac{N L}{ ||\Psi_0||^2} 
\int \frac{dk}{2 \pi}  \tilde p_0(N k) 
\sum_\lambda \frac{1}{||\Psi_\lambda||^2}
 \langle \Psi_0 | \hat \rho(0) | \Psi_\lambda \rangle 
\langle \Psi_\lambda | \hat \rho(0) | \Psi_k \rangle 
e^{ i N k y + i K_\lambda (x-y) } e^{- (E_\lambda-E_0) t} 
\eea
where we used that $\langle \Psi_\lambda | \hat \rho(y) | \Psi_k \rangle = 
e^{ i (N k - K_\lambda) y } \langle \Psi_\lambda | \hat \rho(0) | \Psi_k \rangle$. 
\\

We will only give here the contribution of the ground state manifold
\bea
\langle \rho(x,t) \rho(y,0) \rangle|_{\rm N-string} =   
\int \frac{dK}{2 \pi}  \tilde p_0(K) 
\int \frac{dP}{2 \pi} 
\frac{e^{i K y + i P (x-y) - \frac{1}{N} P^2 t} }{\prod_{a=1}^{N-1} (1 + (\frac{P}{N \bar c a} )^2) \prod_{a=1}^{N-1} (1 + (\frac{K-P}{N \bar c a} )^2)}
\eea
For any $N$ it can be rewritten as as a superposition of packets of particles with a random center
\be \label{avinit} 
\langle \rho(x,t) \rho(y,0) \rangle|_{\rm N-string} =
\mathbb{E}_\omega \left( \int dx_s \, p_0(x_s) \rho_N^s(y-x_s) \rho_N^s(x-x_s-\sqrt{\frac{2 t}{N}} \omega)  \right)
\ee
where $\omega$ is a unit Gaussian random variable. This
shows that the ground state manifold (i.e. the $N$-string contribution) captures only the
diffusion of the center of mass, i.e. the position of the shock, averaged over its initial position.
In the large $N$ limit one can neglect that diffusion, which corresponds to 
neglecting the noise in the Burgers equation. The ground state manifold then only
captures the correlation due to the random initial position of the shock. 
\\

Note that the Burgers dynamics in the large $N$ limit contains much more information on time scales of order unity. It is the information contained in the excited states of the LL whose energy above the ground state scale as $\sim N^2 \bar c^2= \bar \gamma^2$ and remains finite in that limit (and are larger or equal to the gap $\gamma_N \simeq 
\frac{1}{4} \bar c^2 N^2 = \frac{1}{4} \bar \gamma^2$). 

\subsection{Limitations of the mapping}

To investigate the decay rate to the stationary state of the RD system, in this subsection we 
solve directly the case $N=2$ in the repulsive case $\bar c>0$, setting $T=1$. The Langevin 
equation reads 
\bea 
&& \dot x_1 = \bar c \, {\rm sgn}(x_2-x_1) + \sqrt{2} \xi_1(t) \quad , \quad 
  \dot x_2 = \bar c \, {\rm sgn}(x_1-x_2) + \sqrt{2} \xi_2(t)
\eea 
Let us denote $x(t)=\frac{1}{2} (x_1(t) + x_2(t))$ and $y(t)=x_2(t)-x_1(t)$. While the center of mass
performs an independent unit Brownian motion, $\dot x(t) = \xi(t)$, the relative coordinate evolves as
\be 
\dot y = - 2 \bar c \, {\rm sgn}(y) + 2 \eta(t)
\ee 
where $\eta(t)$ is an independent unit white noise. 
Its probability density $P(y,t)$ then evolves according to
\be  \label{PP} 
\partial_t P = 2 \partial_y^2 P + 2 \bar c \partial_y ({\rm sgn}(y)  P)
\ee 
Introducing the Laplace transform $\tilde P(y,s)= \int_0^{+\infty} dt P(y,t)$ one obtains
\be  \label{equas} 
s \tilde P - P(y,0) = 2 \partial_y^2 \tilde P + 2 \bar c \partial_y ({\rm sgn}(y)  \tilde P)
\ee 
where $P(y,0)$ is the initial condition.
\\

Let us choose a simple (even) initial condition, $P(y,0)= \frac{a}{2} e^{- a |y|}$ where $a>0$. 
It is then easy to solve separately for $y>0$ and $y<0$. There are two integration constants on both sides.
One on each side is set to zero by requiring that $\tilde P(y,s) \to 0$ at $y \to \pm \infty$ for $s>0$. 
Continuity of $P$ at $y=0$ gives another condition and finally, from integrating \eqref{PP} on a small
interval around $y-0$ one obtains the matching condition 
\be  \label{matching} 
P'(0^+,t) - P'(0^-,t) + 2 \bar c P(0,t) = 0 
\ee 
and the same relation holds for the Laplace transforms.
\\

For simplicity of notations let us choose units so that $\bar c=1$. This leads to the unique solution 
\be 
\tilde P(y,s) = \frac{1}{s + 2 a(1- a)} \left( \frac{a}{2} e^{- a |y|} + \frac{a(1- a)}{\sqrt{1+ 2 s} -1} e^{- \frac{1}{2} (1+ \sqrt{1+ 2 s}) |y|} \right) 
\ee 
The first term is a special solution of \eqref{equas}, in real time
\be
P_s(y,t) = \frac{a}{2} e^{- a |y|} e^{- 2 a (1-a) t} 
\ee 
which however does not obey the matching condition \eqref{matching}, except for $a=1$ in which case
it is the stationary solution
\be
P_{\rm stat} (y) = \frac{1}{2} e^{- |y|} 
\ee 

Returning to the case of general $a$ let us invert the Laplace transform at $y=0$. One finds
\be 
P(0,t) =
\frac{1}{4} \left((2 a-1) e^{2 (a-1) a t}
   \left(\text{erf}\left(\frac{(1-2 a)
   \sqrt{t}}{\sqrt{2}}\right)+1\right)+\text{erf}\left
   (\frac{\sqrt{t}}{\sqrt{2}}\right)+1\right)
\ee

There is a transition in the large time behavior at $a=1/2$. For $a \neq 1/2$ one finds as $t \to +\infty$
\be 
P(0,t) =  \frac{1}{2} + e^{-\frac{t}{2}} \left(\frac{a (a-1) \sqrt{2} }{(1-2 a)^2 \sqrt{\pi } t^{3/2}}+O(t^{-5/2})\right) - \frac{1}{2} (1-2 a) e^{- 2 a (1- a) t}   \, \theta(a<\frac{1}{2}) 
\ee 
where the asymptotic value is $P_{\rm stat} (0)=1/2$. For $a>1/2$ this decay is consistent with the result from the LL model
with decay rate $\gamma_2= \frac{1}{2}$. For $a<1/2$ there is an additional contribution with a slower decay rate $2 a (1- a) \leq \frac{1}{2}$. At the transition point $a=1/2$ one finds
\be 
P(0,t) =  \frac{1}{4}
   \left(\text{erf}\left(\frac{\sqrt{t}}{\sqrt{2}}\right)+1\right) = \frac{1}{2} +
   e^{-\frac{t}{2}}
   \left(-\frac{1}{2 \sqrt{2 \pi t
   }}+ \frac{1}{2 \sqrt{2 \pi} t^{3/2}}+O( t^{-5/2}) \right) 
\ee 
\\

For completeness we gives also $P(y,t)$ from Laplace inversion. One finds for any $a$
 \bea \label{Pyt} 
 && P(y,t) =  \frac{1}{2} a e^{-2
   (1-a) a t-a |y|}
\\
&& 
   \frac{1}{4} e^{-|y|} \left(-a e^{(a-1) (2 a t-|y|)}
   \text{erfc}\left(\frac{-4 a t+2 t+|y|}{2 \sqrt{2}
   \sqrt{t}}\right)+(a-1) e^{a (2 (a-1) t+|y|)}
   \text{erfc}\left(\frac{4 a t-2 t+|y|}{2 \sqrt{2}
   \sqrt{t}}\right)+\text{erfc}\left(\frac{|y|-2 t}{2
   \sqrt{2} \sqrt{t}}\right)\right) \nonumber   \eea 
 
 At the transition point $a=1/2$ it simplifies into
\bea P(y,t) =
\frac{1}{4} e^{-|y|} \left(e^{\frac{|y|-t}{2}}
   \text{erf}\left(\frac{|y|}{2 \sqrt{2}
   \sqrt{t}}\right)+\text{erfc}\left(\frac{|y|-2 t}{2
   \sqrt{2} \sqrt{t}}\right)\right)
\eea 
Interestingly we see that \eqref{Pyt} at fixed $y,t$ is nicely analytic in $a$ across $a=1/2$. Hence there is a way to
extract the result in the regime $a<1/2$ from the LL, presumably by analytic continuation of the overlap integrals
(i.e. by moving the contours of integration on the string momenta $k_j$). We leave that study to the future. 
\\

{\it General initial condition}. Note that the general even initial condition can be solved by introducing $\hat P(\mu,s)= \int_0^{+\infty} dy e^{- \mu y} \tilde P(y,s)$. 
Denoting $P_0(\mu)= \int_0^{+\infty} dy e^{- \mu y} P(y,0)$, the Laplace transform of the initial
condition, one obtains
\be
s \hat P(\mu,s) - P_0(\mu) = 2 \mu (1+ \mu) \hat P(\mu,s) - 2 \tilde P'(0,s) - 2 (1+ \mu) P(0,s)
\ee 
and we want to impose $P'(0,s)+P(0,s)=0$. The solution is
\be 
\hat P(\mu,s)= \frac{P_0(\mu) - 2 \mu \tilde P(0,s)}{s - 2 \mu(1+ \mu)} 
\ee 
To determine the unknown function $\tilde P(0,s)$ one notes that there should not be a pole at $s=2 \mu(1+\mu)$. Indeed since $\mu>0$ that would lead to a Laplace inverse diverging exponentially with time. Hence one has
\be 
\tilde P(0,s= 2 \mu(1+\mu)) = \frac{P_0(\mu)}{2 \mu} 
\ee 
Equivalently, setting $\mu= \frac{1}{2} (\sqrt{1+2 s}-1)$ (the positive root) one must have
\be 
\tilde P(0,s) = \frac{P_0(\mu=\frac{1}{2} (\sqrt{1+2 s}-1))}{\sqrt{1+2 s}-1} 
\ee 
In the case of the above initial condition $P_0(\mu)=\frac{a}{2} \frac{a}{a+\mu}$, Laplace inverting 
in $\mu$ one recovers the above results. In that case the large time decay rate can be obtained
as $- 2 \mu(1+\mu)$ with $\mu=-a$ the pole of $P_0(\mu)$. Presumably this is a general feature,
with $\mu$ being the smallest singularity of the Laplace transform of the initial condition.

\subsubsection{LL calculation}

Let us compare with the prediction from the LL model for $N=2$. We set 
$\bar c=1$. We recall that $\Psi_0(\vec x)= 2 e^{- \frac{1}{2} |x_2-x_1|}$. 
The initial condition is
\be 
P_2(x_1,x_2,t)= \frac{a}{2} e^{- a |y|} p_0(x) 
\ee 
with $y=x_2-x_1$ and $x=\frac{x_1+x_2}{2}$. The contribution of the $2$-string is simply
\be
P|_{n_s=1}(x,y,t)= \frac{1}{2} e^{- |y|} p_t(x) \quad , \quad p_t(x)=\int \frac{dK}{2 \pi} e^{i K x - \frac{1}{2} K^2 t} \tilde p_0(K) 
\ee 
where $\tilde p_0(K)= \int dx e^{- i K x} p_0(x)$. The overlap integral with the ground state always exists, and
does not generate any condition on the decay of the initial PDF of $y$.

The contribution of the two $1$-strings is, with the definitions $k_1=k-q/2$, $k_2=k+q/2$ 
\be 
P|_{n_s=2}(x,y,t) = \frac{1}{2} e^{- \frac{1}{2} |y|}  e^{-t/2}  \int \frac{dk}{2 \pi} \frac{dq}{2 \pi} \frac{q^2}{q^2 + 1} e^{-2 k^2 t - \frac{q^2}{2} t} 
\Psi_{k_1,k_2,1,1}(x,y) \langle \Psi(0) | \Psi_{k_1,k_2,1,1} \rangle
\ee 
where 
\be 
\Psi_{k_1,k_2,1,1}(x,y) = e^{2 i k x} \phi_q(y) \quad , \quad \phi_q(y)=  e^{ \frac{1}{2} i q |y|} (1 + \frac{i}{q}) 
+ e^{ - \frac{1}{2} i q |y|} (1 - \frac{i}{q}) 
\ee 

The overlap exists, strictly speaking, only for $a > 1/2$ and is then equal to 
\bea 
&& \langle \Psi(0) | \Psi_{k_1,k_2,1,1} \rangle = \int d^N\vec x  \frac{P_2(\vec x,t=0)}{\Psi_0(\vec x)} \Psi^*_{k_1,k_2,1,1}(\vec x)  = [ \int dx' e^{-2 i k x'} p_0(x') ]  \times \int dy' \frac{a}{4} e^{-(a-\frac{1}{2}) |y|} \phi^*_q(y)  \\
&& =  \tilde p_0(2 k) \frac{4 a(a-1)}{q^2 + (2 a-1)^2}
\eea 
Thus we obtain
\be 
P|_{n_s=2}(x,y,t) = P|_{n_s=2}(y,t) p_t(x) 
 ~,~ P|_{n_s=2}(y,t)  =
 a (a-1) e^{- \frac{1}{2} |y|-t/2}  
\int \frac{dq}{2 \pi} \frac{1}{q^2 + (2 a-1)^2}  \frac{q^2 e^{- \frac{q^2}{2} t} }{q^2 + 1} ( e^{ \frac{1}{2} i q |y|} (1 + \frac{i}{q}) + c.c) 
\ee 
To compute these integrals we use 
\be 
e^{- \frac{q^2}{2} t}  = \frac{1}{\sqrt{2 \pi}} e^{ i q u \sqrt{t} - u^2/2} 
\ee 
The integral over $q$ can then be easily obtained from mathematica Fourier transform routine and 
the result integrates explicitly over $u$ in terms of error functions. Putting both contributions
together one obtains exactly the result \eqref{Pyt} obtained by the direct method, although here
its derivation is valid only for $a>1/2$.

\section{Large $N$ at fixed $\bar c$: inviscid Burgers dynamics}
\label{sec:fixed} 

As in the text we first consider the limit of large $N$ at fixed $\bar c$, 
and scale $V(x)=N \tilde V(x)$ and rescale time as $t=\tau/N$. 
In this limit the dynamics is described by the equation \eqref{1storder} in the text. This first-order PDE can exhibit two common types of singularities (recalling that $\partial_x r \geq 0$):

(i)  intervals $[x^-(\tau),x^+(\tau)]$ empty of particles, where $r(x,\tau)=r_p$ has a plateau 
(i.e. $\rho(x,\tau)=0$). The reciprocal function $x(r,\tau)$ (used below) has a jump at $r=r_p$.
Note that the value of the plateau $r_p$ is time independent (as long as the plateau exists).

(ii) shocks at $x=x_s(\tau)$ where $r$ exhibits a upward jump $r(x_s^\pm,\tau)=r^\pm(\tau)$
where a macroscopic fraction $r^+(\tau)-r^-(\tau)>0$ of particles accumulate and the density
has a delta peak. There the
reciprocal function $x(r,\tau)$ (used below) has a plateau on $[r^-(\tau),r^+(\tau)]$

While empty regions can exist for both signs of $c$, shocks occurs in the dynamics when $\bar c<0$.

\subsection{Repulsive interactions $c=- \bar c>0$} 

Let us write in that case with $c>0$
\bea \label{1storder2} 
\partial_\tau r(x,\tau) = \left( \tilde V'(x) - 2 c  ~ r(x,\tau) \right) \partial_x r(x,\tau)
\eea 
To interpret the term in factor let us note that from the definition \eqref{defrank} of the rank field the original Langevin equation \eqref{langevin1} can be rewritten in these rescaled variables as
\be  \label{langevinnew} 
\dot x_i= \frac{d}{d\tau} x_i(\tau) =  2 c \, r(x_i(\tau),\tau) - \tilde V'(x_i(\tau)) + \frac{\sqrt{2 T}}{\sqrt{N}} \tilde \xi_i(\tau)
\ee 
where $\tilde \xi(\tau)$ is a unit white noise. Thus neglecting the noise term at large $N$ one sees that 
the prefactor in \eqref{1storder2} is simply $ - \dot x$.


Setting $\partial_\tau r=0$ in \eqref{1storder} we see that around a given $x$ a 
stationary solution $r_{\rm stat} (x)$ is either constant, with $r_{\rm stat} '(x)=0$, e.g. part of an empty interval, 
or equal to $r_{\rm stat} (x)=\frac{1}{2} \tilde V'(x)/c$,
which is acceptable only if $\tilde V''(x) \geq 0$. 
\\

\begin{figure}[ht]
\includegraphics[angle=0,width=0.5\linewidth]{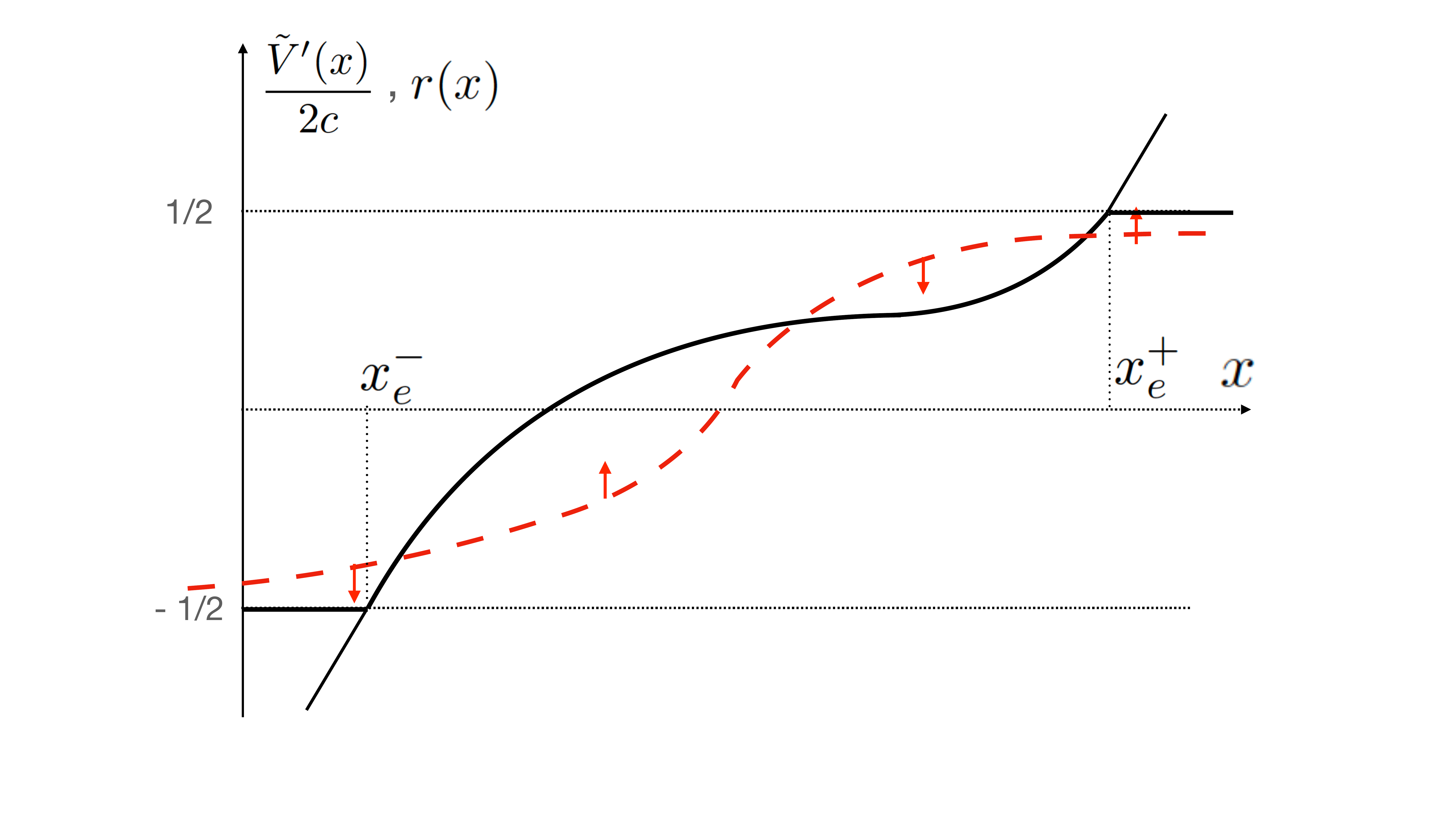}
\caption{Light black: plot of $\tilde V'(x)/(2 c)$ versus $x$, convex potential case. Thick black: stationary rank function $r_{\rm stat} (x)$. The support of the density $r'_{\rm stat} (x)$ is a single interval $[x_e^-,x_e^+]$. Dashed red: initial $r_0(x)$. Red arrows indicate the variation of $r(x,t)$ with time ($\dot r$ has opposite sign to $\dot x$)}
\label{Fig:Rep1}
\end{figure}

\begin{figure}[ht]
\includegraphics[angle=0,width=0.5\linewidth]{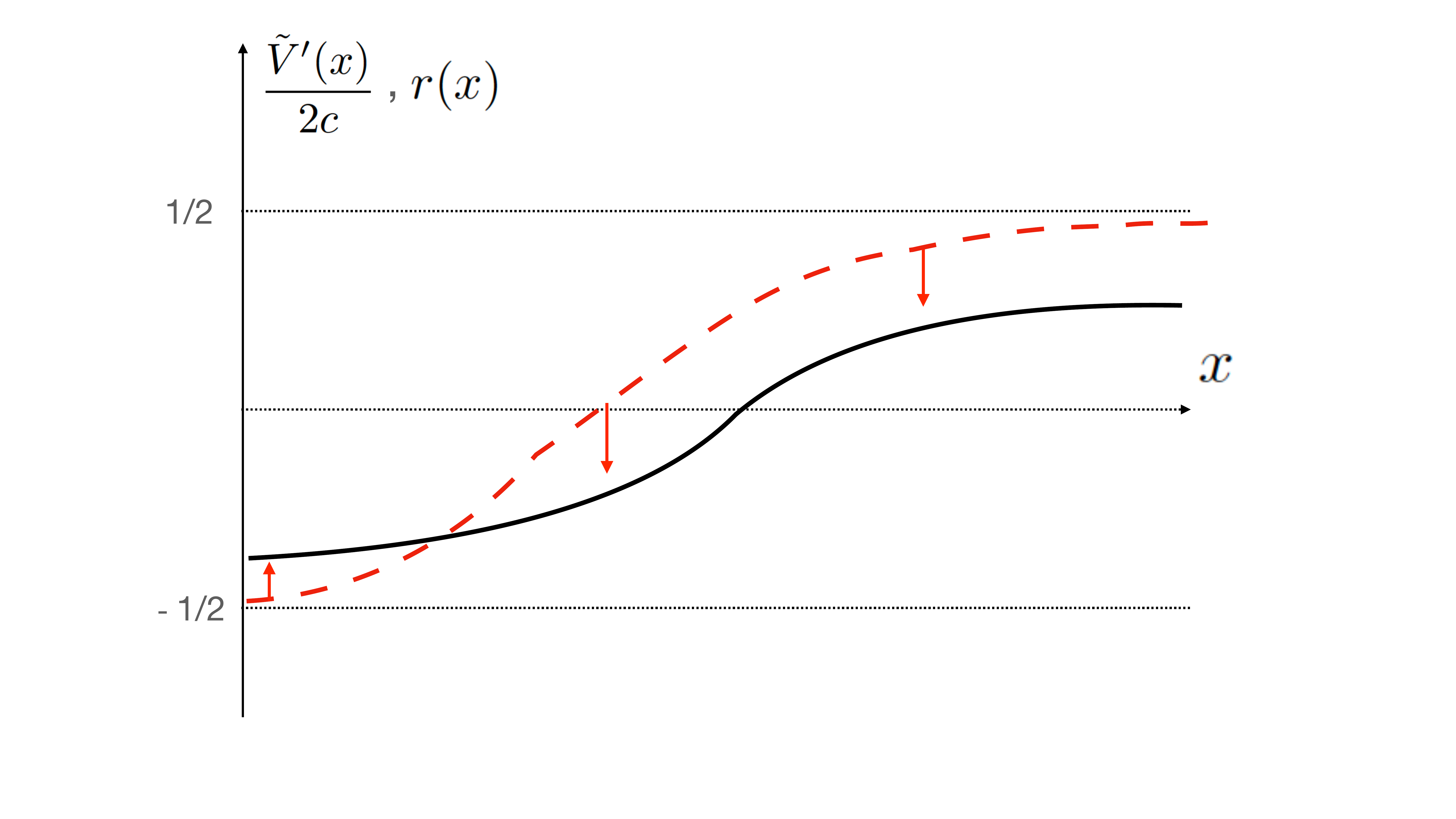}
\caption{Light black: plot of $\tilde V'(x)/(2 c)$ versus $x$. Convex potential but not confining enough so that 
some particles escape to infinity. Thick black: stationary $r_{\rm stat} (x)$, its support is the real axis. Dashed red: initial $r_0(x)$. Red arrows indicate the variation of $r(x,t)$ with time $\dot r$ has opposite sign to $\dot x$}
\label{Fig:Rep3}
\end{figure}

{\it Convex potential}. The simplest case is when $\tilde V(x)$ is convex. There are two subcases.
Suppose first that $\tilde V(x)$ is sufficiently confining, i.e. $\tilde V'(+\infty)>c$ 
and $\tilde V'(-\infty)<-c$.  This is shown in Fig. \ref{Fig:Rep1}. In that case 
the stationary solution is unique, and the density is supported on a finite interval
$[x_e^-,x_e^+]$ where the two edges $x_e^\pm$ are the roots of $\tilde V'(x_e^\pm)=\pm c$.
On this interval the rank field and the density are given by
\be \label{aga} 
r_{\rm stat} (x)= \frac{\tilde V'(x)}{2 c}  \quad , \quad   \rho_{\rm stat} (x)=\frac{\tilde V''(x)}{2 c}  \quad , \quad
x_e^- < x < x_e^+ 
\ee 
Outside of this interval one has $r_{\rm stat} (x)=\frac{1}{2} {\rm sgn}(x)$ and the density vanishes.
The density thus exhibits generically a jump at these edges. 

What is the dynamics toward stationarity? From \eqref{1storder} we see that $r(x,\tau)$ increases with $\tau$ in regions such that $r(x,\tau) < \tilde V'(x)/2c$ (i.e. $\dot x<0$ in \eqref{langevinnew} and the particles move to the left) 
and decreases if $r(x,\tau)>\tilde V'(x)/2c$ (i.e. $\dot x>0$ and the particles move to the right). This is illustrated in  Fig. \ref{Fig:Rep1} and in that case of the convex potential it leads to the
convergence to the unique stationary solution, Eq \eqref{aga}. 

There is a second subcase however, when either $\tilde V'(+\infty)<c$ or $\tilde V'(-\infty)>-c$ or both.
It is illustrated in Fig. \ref{Fig:Rep3}. The stationary measure is again unique and 
given by Eq. \eqref{aga} for all $x$, i.e. the edges $x_e^\pm$ are pushed to infinity. In that case, a finite fraction of the particles are expelled to $\pm \infty$: a fraction $\max(0,\frac{1}{2} - \frac{\tilde V'(+\infty)}{2 c})$ to $+\infty$
and a fraction $\max(0,\frac{\tilde V'(+\infty)}{2 c}+ \frac{1}{2})$ to $-\infty$.
\\

\begin{figure}[ht]
\includegraphics[angle=0,width=0.5\linewidth]{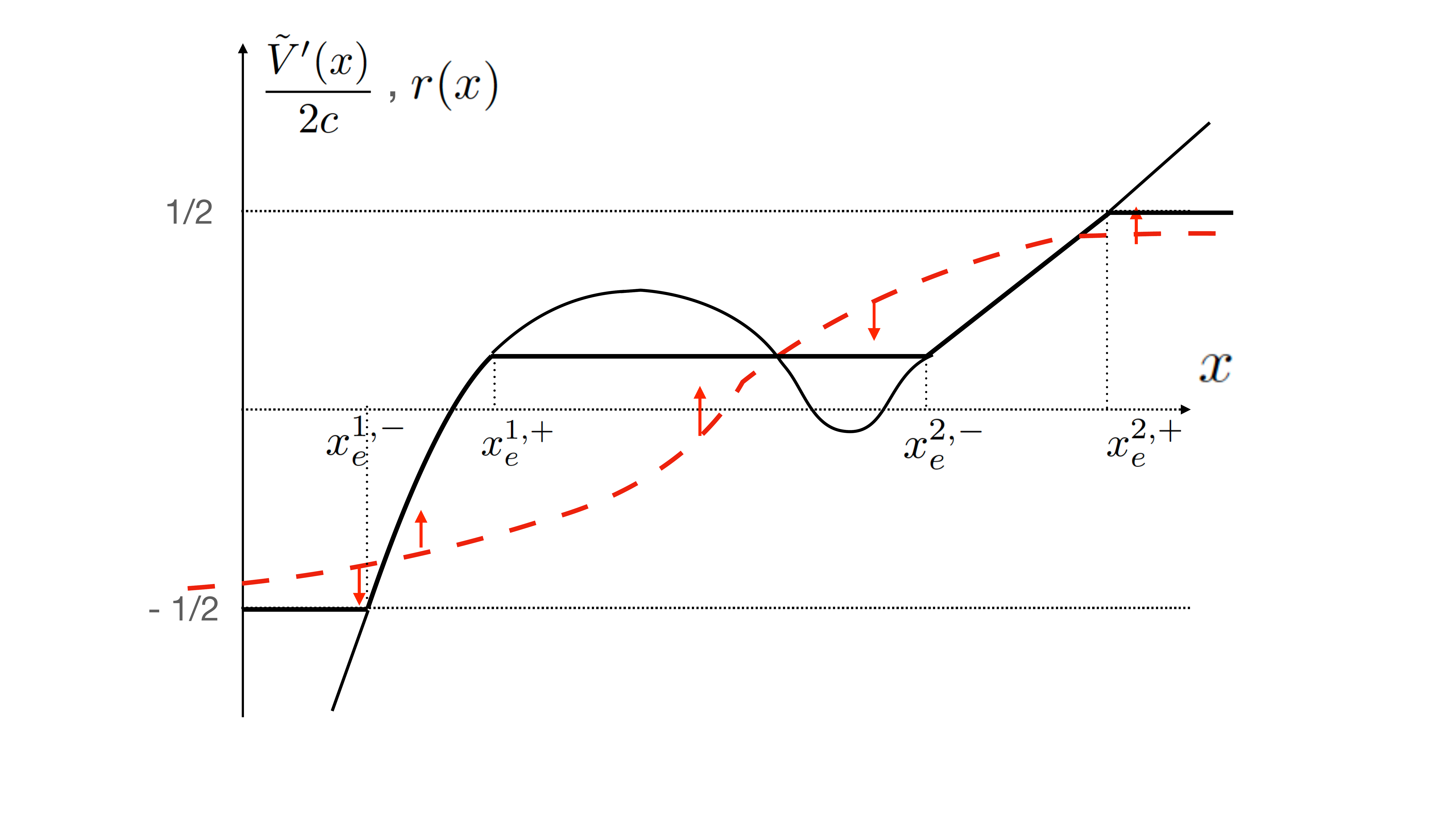}
\caption{Light black: plot of $\tilde V'(x)/(2 c)$ versus $x$ for a double well potential. Thick black: stationary $r_{\rm stat} (x)$.
The support of the density $r'_{\rm stat} (x)$ consists of two intervals. Dashed red: initial $r_0(x)$. Red arrows indicate the variation of $r(x,t)$ with time $\dot r$ has opposite sign to $\dot x$. The position of the plateau in $r_{\rm stat} (x)$
is determined by $r_0(a)$ where $a$ is the intersection of $r_0(x)$ and $\tilde V'(x)/(2 c)$ in the region where
the potential is concave.}
\label{Fig:Rep2}
\end{figure}

{\it Multiple wells}. For double (or multiple) well types potentials the situation is more involved, since there
are regions with $\tilde V''(x) <0$ where \eqref{aga} cannot hold. There are thus families of stationary states
which depend on the initial condition $r_0(x)=r(x,0)$. They consist in a sequence of intervals where \eqref{aga} 
holds, separated by empty regions. An example is shown in Fig. \ref{Fig:Rep2} for a given initial condition.
In that example at large time the support of the density consists of two intervals $[x_e^{j,-},x_e^{j,+}]$, $j=1,2$.
The outer edges are still given by the roots $\tilde V'(x_e^{1,-})=-c$ and $\tilde V'(x_e^{2,+})=-c$. 
The inner edges $x_e^{1,+}$ and $x_e^{2,-}$ can be found using the following property. 
Any point $x=a$ such that $r_0(a)=\tilde V'(a)/2 c$ has $\partial_\tau r(a,\tau)=0$ hence $r(a,\tau)=r_0(a)$ 
at all times. This is true assuming that $\partial_x r(x,\tau)|_{x=a} < +\infty$, which holds for the repulsive case
(it fails for the attractive case, see below). The point $x=a$ is a fixed point $\dot x=0$ of the flow 
\eqref{langevinnew}. There are thus two cases 

(i) $\tilde V''(a) - 2 c r'_0(a) >0$ in which case $\frac{d\dot x}{dx}|_{x=a} <0$  and $x=a$ is an attractive fixed point for the flow
(such as $x_1$, $x_3$ in Fig. \ref{Fig:Rep2}) which thus belongs to the support of the stationary state.

(ii) $\tilde V''(a) - 2 c r'_0(a) <0$ in which case $\frac{d\dot x}{dx}|_{x=a} >0$  and $x=a$ is a repulsive fixed point for the flow. If $\tilde V''(a)>0$, as for $x_2$ in Fig. \ref{Fig:Rep1}), the point $a$ belongs to the support of the stationary state.
However if $\tilde V''(a)<0$, such as $x_2$ in Fig. \ref{Fig:Rep2}), it cannot belong to the support. Then in the large time limit, $r(x,\tau)$ develops a plateau around the point $x=a$, 
of value $r_\infty(x)=r_0(a)=V'(a)/(2 c)$. This determines uniquely the asymptotic state from the initial condition, within the family of possible stationary states, see Fig. \ref{Fig:Rep2}. To put it simply, all the particles to the left of $x=a$ end up in the first well, and all those to the right of $a$ end up in the second well. 

Note that the sign of $V''(a) - 2 c \partial_x r(x,\tau)|_{x=a}$ does not change with time since
\be \label{eqdiff} 
\partial_\tau (V''(a) - 2 c \partial_x r(x,\tau)|_{x=a}) = -2 c \partial_x \partial_\tau  r(x,\tau)|_{x=a}
= -2 c  (V''(a) - 2 c \partial_x r(x,\tau)|_{x=a}) \partial_x r(x,\tau)|_{x=a}
\ee 
Using \eqref{1storder2} and that $(V'(a)- 2 c r(a,\tau) ) \partial_x^2 r(x,\tau)|_{x=a} = 0$ 
(if we assume $\partial_x^2 r(x,\tau)|_{x=a} < +\infty$ which holds in the repulsive case). 
\\

{\it General solution for the dynamics}. It is easy to obtain the general solution of \eqref{1storder} by considering $\tau(x,r)$ and
using $\frac{\partial_x r}{\partial_\tau r}= - \partial_x \tau$. We obtain 
$(\tilde V'(x)-2 c r)\partial_x \tau=-1$. Let us denote $x_0(r)$ the inverse
function of the initial condition $r_0(x)=r(x,0)$, i.e. $r_0(x_0(r))=r$.
The general solution is then for $r \in [-\frac{1}{2},\frac{1}{2}]$
\be \label{solugen1} 
\tau=\int_{x_0(r)}^x \frac{dy}{2 c r - \tilde V'(y)} 
\ee 
\\

{\it Dynamics without potential}. Consider first the system in the absence of external potential, $\tilde V(x)=0$. Eq. \eqref{solugen1}  
recovers the perturbative solution of the inviscid Burgers equation 
\be \label{eqeq0} 
2 c r \tau = x-x_0(r) \quad , \quad r = r_0(x - 2 c r \tau) 
\ee
equivalently $r(x,\tau) = r_0(w(x,\tau))$ where
\bea \label{eqeq} 
w + 2 c ~ r_0(w) \tau = x  \quad  \Leftrightarrow \quad w=w(x,\tau) 
\eea 
This solution is valid as long as the map $w \to w + 2 c ~ r_0(w) \tau$ is 
invertible, i.e. it is valid for all times $\tau$ such that the initial density satisfies
\bea \label{cond} 
1 + 2 c \tau \rho_0(x) >0 \quad \forall x
\eea
In the repulsive case, $c>0$, it is thus valid for all times. Since $\tilde V(x)=0$ the 
particles are expelled to infinity, and for a localized initial condition \eqref{eqeq0} 
leads to $r(x,\tau) \simeq \frac{x}{2 c \tau}$ for $|x| < 2 c \tau$
at large $\tau$, a non-stationary solution which has the form of 
a symmetric front moving at speed $\pm 2 c$. 
An instructive example is to consider a square initial condition $\rho_0(x)=\rho(x,0)=\frac{1}{2\ell} \theta(\ell-|x|)$
with $\ell>0$.
The exact solution is then 
\bea \label{square1} 
r(x,\tau) = \begin{cases} \frac{x}{2( \ell + c \tau)} \quad , \quad |x| < \ell + c \tau \\
\pm 1/2 \quad , \quad ~~\, |x| > \ell + c \tau \end{cases}
\eea 
hence the density remains a square at all times, with a width growing linearly with time
\bea \label{solurep}
\rho(x,\tau) = \frac{1}{2 (\ell+ c \tau)} \theta(\ell + c \tau-|x|)
\eea 
i.e. the gas expands ballistically.

In general one can check that the perturbative solution satisfies both that (i) the point $w=a$ such that 
$r_0(a)=0$ is preserved, since at time $\tau$ it corresponds from \eqref{eqeq} to $x=a$, hence 
$r(a,\tau)=r_0(a)=0$, and (ii) the center of mass does not move since its position at any time is
$\int_{-1/2}^{1/2} dr x = \int_{-1/2}^{1/2}  
dr_0 (w + 2 c \tau r_0) = \int_{-1/2}^{1/2} dr_0 w$ using that $[r_0^2]_{-1/2}^{1/2}=0$. 
Thus the gas expands both such that the number of particle to the left and to the right of $x=a$ remains the same and so that the center of mass of the gas is preserved. 
\\

{\it Harmonic well}. For the harmonic well, $\tilde V'(y)=\mu_0 y$ the solution $r=r(x,\tau)$ from
\eqref{solugen1} 
takes the implicit form
\be
e^{- \mu_0 \tau}= \frac{2 c r-\mu_0 x}{2 c r - \mu_0 x_0(r)} 
\ee 
Consider for instance the case where all particles start at $x=y$, i.e.
$x_0(r)=y$. One finds that the density is uniform 
\be \label{soluquadratic} 
r(x,\tau) = \frac{\mu_0}{2 c} \frac{x-y e^{-\mu_0 \tau}}{1- e^{- \mu_0 \tau} }
~,~ \rho(x,\tau)= \frac{\mu_0}{2 c(1- e^{- \mu_0 \tau})}
\ee
in the time dependent interval $x \in [x_e^-(\tau),x_e^+(\tau)]$ with $x_e^\pm(\tau) = \pm \frac{c}{\mu_0} (1- e^{- \mu_0 \tau}) + y e^{-\mu_0 \tau}$, and zero outside. It converges exponentially fast to the stationary state
\bea \label{dens} 
 \rho_{\rm stat} (x) = \begin{cases}  \frac{\mu_0}{2 c}    \quad , \quad |x| < \frac{c}{\mu_0} \\ 0 \quad , 
 \quad |x| > \frac{c}{\mu_0} \end{cases}
\eea 
i.e. a constant density in a finite interval, as was found in the Jellium studies in the regime
large $N$ at fixed $c=-\bar c$. 
\\

{\it Inverse harmonic well}. Note that the solution \eqref{soluquadratic} also holds for $\mu_0<0$ (concave potential).
In that case the gas expands exponentially fast, the two edges being given by $x_e^\pm(\tau) =  (y \pm \frac{c}{|\mu_0|}) e^{|\mu_0| \tau} \mp \frac{c}{\mu_0}$. The density inside the interval vanishes exponentially fast
$\rho(x,\tau)= \frac{|\mu_0|}{2 c(e^{|\mu_0| \tau}-1)}$ an example of the formation of a plateau 
in a region with $\tilde V''(x) <0$, as discussed above.

\subsection{Attractive interactions $\bar c=-c>0$}

In the case of attractive interactions some of the above formula are still valid, but only up to some
finite time at which there is formation of delta peaks in the density, which, in the scaling considered here, 
correspond to packets of particles containing a finite fraction of the $N$ particles. In the context of the
Burgers equation for $r(x,t)$ it corresponds to the formation of shocks. In the absence of external potential 
it corresponds to the string solutions of the LL model. 
\\

Consider the case $\tilde V(x)=0$. The simplest example is
the exact solution \eqref{solurep} 
\bea \label{solurep}
\rho(x,\tau) = \frac{1}{2 (\ell- \bar c \tau)} \theta(\ell - \bar c \tau-|x|)
\eea 
The gas now contracts ballistically. The density becomes a delta function at $x=0$,
 i.e. $r(x,\tau)$ develops a jump (i.e. a shock) of unit amplitude at $x=0$, 
 in a finite time $\tau=\ell/\bar c$.
 
In fact the solution \eqref{solurep} is very special, as the final position of the shock is
its "naive" position. This is given by arguments similar to those of the previous subsection,
i.e. it is the attractive point of the dynamics \eqref{langevinnew}, equal to
the root $x=a$ of $- \dot x \equiv \tilde V'(x) + 2 \bar c r_0(x)=0$, in that case $a=0$ (since $\tilde V(x)=0$).

More generally, when $\rho(x,0)$ is not uniform, the perturbative solution of the Burgers equation 
\eqref{eqeq0}, \eqref{eqeq} breaks down before, and from \eqref{cond}, a shock appears at time
\bea
\tau=\tau_s = \frac{1}{2 \bar c \max_x \rho(x,0)}
\eea 
and at position $x=x_s= {\rm argmax} \rho(x,0)$, the location of the maximum of the density.
The dynamics beyond that time is more complex (especially in presence of a potential)
and one cannot use the above "naive" argument
to predict the position of the shock at large time. This argument to predict the final state works
only in the repulsive case, when the perturbative solution holds for all times. 
\\

To describe fully the dynamics in the case of attractive interactions, including shocks,
one must give a more precise meaning to the inviscid Burgers equation \eqref{1storder2} 
allowing for shocks. This is a standard problem, and the solution can be deduced from
the inclusion of the (very small) neglected diffusion term. Anticipating on the study of this term in
the next section (note that the noise term remains
subdominant as compared to the diffusion term),
the proper solution in the case $\tilde V(x)=0$ to which we restrict here, is 
\bea
&& r(x,\tau) = \frac{w(x,\tau)-x}{2 \tau \bar c} = r(w(x,\tau),0) \\
&& w(x,\tau) := {\rm argmin}_{w \in \mathbb{R}} [ E_{x,\tau}(w)  ] \quad , \quad 
E_{x,\tau}(w) = \frac{(w-x)^2}{4 \tau} - \bar c \int_0^w dx' r_0(x')  \label{minimiz} 
\eea
This formula is valid for any sign of $c=-\bar c$. It recovers the perturbative solution of Burgers equation 
when there is a single minimum in \eqref{minimiz}, which holds for all
times in the repulsive case (since $E_{x,\tau}(w)$ is convex in that case).
In the attractive case, the potential energy $-\bar c \int_0^w dx' r_0(x')$ is a concave function
with linear behavior at infinity, $\simeq \frac{\bar c}{2} |w|$. Beyond the time $\tau=\tau_s$, $E_{x,\tau}(w)$ has a (time dependent) number $n+1$ 
of local minima $w_i=w_i(x,\tau)$, $w_1<\dots<w_n$, solutions of
\be  \label{locmin} 
\frac{w_i- x}{ 2 \bar c \tau} = r_0(w_i)
\ee
and $n$ maxima (each between two minima).
For any given $\tau$ and $x$ the global minimum belongs to this set,
$w(x,\tau)={\rm argmin}_{w \in \{w_i\}}  E_{x,\tau}(w)$. As $x$ is increased (at fixed $\tau$) 
the global minimum $w(x,\tau)$ takes increasing values in a subset of $p \leq n$ elements
$w_{i_1}<\dots<w_{i_p}$ of $\{w_i\}$. 
This corresponds to $p$ shocks, i.e. $p$ packets of particules being present at time $\tau$. 
The positions of the shocks
$x_s^j = x_s^j(\tau)$, 
$x_s^1(\tau) < \dots x_s^p(\tau)$, are determined by the energy degeneracy condition
\be \label{deg} 
E_{x_s^j,\tau}(w_{i_j}) = E_{x_s^j,\tau}(w_{i_{j+1}}) 
\ee 
One has $p \leq n$ because some of the (higher up, small scale) local minima $w_i$ never become the global minimum as $x$ is varied. Accordingly, as $\tau$ increases, the positions of two neighboring shocks may coincide, in which case they merge (i.e. the two delta packets of particles merge), and the intermediate local minima drops from the list. 

A simple graphical construction allows to determine these positions according 
to the "equal area law", see Fig. \ref{Fig:Rep4}, which is a simple consequence of
the degeneracy condition \eqref{deg}.

\begin{figure}[ht]
\includegraphics[angle=0,width=0.5\linewidth]{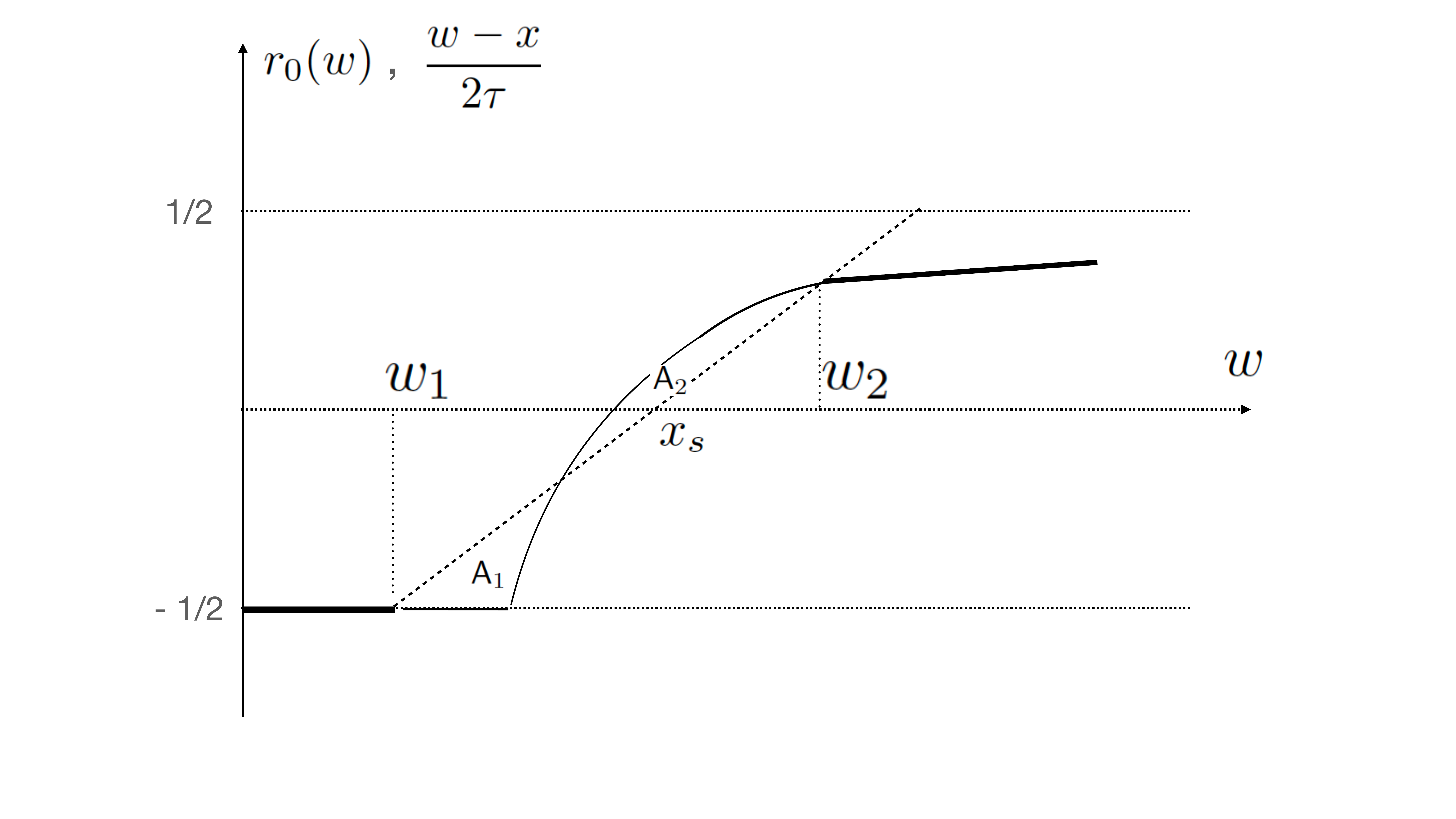}
\caption{Graphical construction of the solution \eqref{minimiz}. Light black: plot of the initial condition
$r_0(w)$ versus $w$. Dashed line: plot of $(w-x)/(2 \tau)$ i.e. a straight line of slope $1/(2 \tau)$ which intercepts
the $w$ axis at $x$. It then intercepts the graph of $r_0(w)$ at $w(x,\tau)$ the minimizer of \eqref{minimiz}.
For some values of $x$ there are three intersections, corresponding
to two minima of \eqref{minimiz} and a maximum. The position of the shock $x=x_s$ is attained
when the areas $A_1$ and $A_2$ are equal. Thick black line: resulting value of $r(x,\tau)=r_0(w)$ 
obtained for $w(x,\tau)<w(x^-,\tau) =w_1$ and for $w(x,\tau)> w(x^-,\tau) =w_2$, which
exhibits a jump at $x=x_s$}
\label{Fig:Rep4}
\end{figure}

At intermediate time there can be a single or several shocks, i.e. delta packets of particles,
on top a a smooth density background. In the large time limit they merge and it always remain only a 
single shock, i.e. a single delta packet containing all the particles, associated to the jump from the leftmost minimum $w_1$ to the rightmost minimum $w_n$. To investigate the late time dynamics let us rewrite the potential in a form convenient 
for the large $|w|$ asymptotics 
\bea  \label{intro} 
&& \int_0^w dx' r_0(x') = \frac{w}{2} - \kappa_+ - g_+(w) = - \frac{w}{2} + \kappa_- + g_-(w)  
\quad , \quad  r_0(w) = \frac{1}{2} - g'_+(w) = - \frac{1}{2} + g'_-(w) 
\eea
where the function $g_+(w)$ decays to zero as $w \to + \infty$,
and $g_-(w)$ decays to zero as $w \to - \infty$. From the local minima condition \eqref{locmin} 
we obtain 
\be \label{w1n} 
w_1 = x -  \bar c \tau + 2 \bar c \tau g'_-(w_1) \quad , \quad w_n = x +  \bar c \tau - 2 \bar c \tau g'_+(w_n) 
\ee 
since $w_1 \to -\infty$ and $w_n \to +\infty$ as $\tau \to +\infty$. The shock position $x_s$ is determined 
by $E_{x_s,\tau}(w_1) = E_{x_s,\tau}(w_n)$ which yields using \eqref{locmin}
\be 
 \bar c^2 \tau (r_0(w_n)^2-r_0(w_1)^2) - \bar c \int_{w_1}^{w_n} dx' r_0(x') = 0
\ee 
which, using \eqref{intro} and its derivative, leads to
\be 
\bar c \tau ( g'_-(w_1) - g'_+(w_n)) (1 - g'_-(w_1) - g'_+(w_n)) = \frac{w_1+w_n}{2} - \kappa_+ - \kappa_- - g_+(w_n)
- g_+(w_1)
\ee 
We now insert \eqref{w1n} and after some simplifications we obtain the shock position as
\be 
x_s = \kappa_+ + \kappa_- + g_+(w_n)
+ g_+(w_1) +  \bar c \tau ( g'_+(w_n)^2 - g'_-(w_1)^2) 
\ee 
Until now this is exact as long as there is only one shock. It can also be written as
\be 
x_s = - \int_{w_1}^{w_n} dx' (r_0(x') - \frac{1}{2} {\rm sgn}(x') ) +  \bar c \tau ( (r_0(w_n)-\frac{1}{2})^2 - 
(r_0(w_1)+\frac{1}{2})^2) 
\ee 
which is again exact and only assumes a single shock with $w_1<0$ and $w_n>0$. 
Inserting the large $\tau$ asymptotics we find $x_s(\tau) = x_s(+\infty) + \delta x_s(\tau)$
where the final position is simply given by the initial center of mass position
\be 
x_s(\infty) = \kappa_+ + \kappa_- = - \int_{-\infty}^{+\infty} (r_0(x') - \frac{1}{2} {\rm sgn}(x') ) 
= \int_{-\infty}^{+\infty} dx x \, \rho(x,0) 
\ee 
and the convergence towards that position can be estimated as
\be 
\delta x(\tau) \simeq g_+(\bar c \tau) + g_-(- \bar c \tau) + \bar c \tau ( g'_+(\bar c \tau)^2 - g'_-(- \bar c \tau)^2) 
\ee 
Similarly the fraction of particles in the shock converges to unity as
\be 
r_0(w_n)-r_0(w_1) = 1 - g'_+(w_n) - g'_-(w_1) \simeq 1 - g'_+(\bar c \tau) - g'_-(-\bar c \tau) 
\ee 
We have assumed that $r_0(x) - \frac{1}{2} {\rm sgn}(x)$ converges to zero faster than $1/|x|$ at large $|x|$, 
equivalently that the center of mass position can be defined. Note that the fact that the center of mass position 
is independent of time when $V(x)=0$ and upon neglecting the noise 
is easily seen by multiplying \eqref{eqrho1} by $x$, integrating over $x$ and using
integrations by part.

\section{Large $N$ with $c=\gamma/N$}

We consider now the large $N$ limit with $\gamma= N c$ fixed, and we also note $\bar \gamma= N \bar c=-\gamma$.
This scaling allows to study in more details the attractive case and the structure of the shocks. 
Here one does not rescale time, nor the potential $V(x)$. 
One sees that in \eqref{eqr} one can still neglect the noise term, but now one must keep the diffusion term, and one obtains ($\gamma$ has arbitrary sign here) the evolution equation of $r(x,t)$ as
\be \label{eqrho23} 
\partial_t r = T  \partial_x^2 r  + 2 \bar \gamma ~ r \partial_x r + V'(x) \partial_x r 
\ee 
where $\gamma$ has arbitrary sign.
\\

\subsection{No external potential} 

Consider first $V(x)=0$. We will look for a solution for $r(x,t)$ in the form
\be \label{rz} 
r(x,t)= \frac{T}{\bar \gamma} \partial_x \log Z(x,t)
\ee 
where $Z(x,t)$ is the solution of the heat equation
\be \label{heat}
\partial_t Z(x,t) = T \partial^2_x Z(x,t) 
\ee 
From \eqref{rz}, we must choose the initial condition to be $Z(x,t=0)= e^{\frac{\bar \gamma}{T} \int_a^x dx' r_0(x')}$.
Here $a$ is arbitrary and we will choose $a=0$ for convenience. Given this initial condition,
if $Z(x,t)$ obeys \eqref{heat} then $H(x,t)=\log Z(x,t)$ obeys $\partial_t H = T \partial_x^2 H + T
(\partial_x H)^2$ and, taking a derivative, $r=\frac{T}{\bar \gamma} \partial_x H$ satisfies 
the Burgers equation \eqref{eqrho23} (with $V(x)=0$), with the proper initial condition. 
Since the solution for $Z(x,t)$ is
\be \label{Z} 
Z(x,t) = \int \frac{dw}{\sqrt{4 \pi T t}} e^{ - \frac{(w-x)^2}{4 T t} + \frac{\bar \gamma}{T} \int_0^w dx' r_0(x') }
\ee 
it implies, using \eqref{rz}, that the solution for $r(x,t)$ is 
\bea \label{soluCH}
r(x,t) 
= \frac{\int \frac{dw}{\sqrt{4 \pi T t}} \frac{w-x}{2 \bar \gamma t} e^{ - \frac{(w-x)^2}{4 T t} + \frac{\bar \gamma}{T} \int_0^w dx' r(x',0) } }{
\int \frac{dw}{\sqrt{4 \pi T t}} e^{ - \frac{(w-x)^2}{4 T t} + \frac{\bar \gamma}{T} \int_0^w dx' r(x',0) }}
\eea 
Note that this solution is valid irrespective of the sign of $\bar \gamma=-\gamma$. 
\\

One can recover the results of Section \ref{sec:fixed} in the fixed $\bar c$ large $N$ limit.
To this aim one sets $t = \tau/N$ and $\bar \gamma = N \bar c$ and take $N \to +\infty$
in \eqref{soluCH}. The argument of the exponential is then uniformly 
of $O(N)$ and from the saddle point method the integral is dominated by the maximum of the integrand.
One then recovers the solution of the inviscid Burgers equation
given above in \eqref{minimiz}. 
\\

\subsubsection{Attractive interactions $\bar \gamma >0$}

Thus, in the large $N$ limit with fixed $\bar \gamma= N \bar c$, if we take $\bar \gamma/T$ of $O(1)$ but large, 
there is a smooth crossover to the results of the previous section. At finite $\bar \gamma>0$ the
shocks have a finite width of order $O(T/\bar \gamma)$ (see below), which vanishes in the limit. 
So here, by keeping the diffusion term, we are looking at the system on the scale of the shock width. 
For a localized initial density there is usually a single shock whose
amplitude is growing. For a multimodal type distribution there may be several shocks
forming which then merge. Each shock contains a finite fraction of the number of
particles. At large time there remains a single shock, which we now study.

One easily checks that for $\bar \gamma>0$ the following initial condition is stationary, $r_{\rm stat} (x)=r(x,t)=r_0(x)$ for all $t$ 
\bea \label{stat2} 
&& \int^w dx' r_{\rm stat} (x')  = \int^w dx' r_0(x')  =  \frac{T}{\bar \gamma} \log[ \cosh[\frac{1}{2 T} \bar \gamma(w -x_s)]] \\
&& r_{\rm stat} (x) = \frac{1}{2} \tanh ( \frac{\bar \gamma}{2 T} (x-x_s) ) \quad , \quad 
\rho_{\rm stat} (x) = \frac{\bar \gamma}{4 T} \frac{1}{\cosh^2(\frac{\bar \gamma}{2 T} (x-x_s) ) }
\eea
either by inserting into \eqref{soluCH} or by checking that it satisfies 
\bea \label{eqrho23} 
T \partial_x^2 r_{\rm stat} (x)  + 2 \bar \gamma ~ r_{\rm stat} (x) \partial_x r_{\rm stat} (x)  = 0
\eea 
It describes a single shock, i.e. a packet containing the $N$ particles. 

The stationary distribution is this not unique since $x_s$ is arbitrary. The low lying diffusion
modes correspond to the diffusion of the center of mass. Indeed it is exact for any $N$ that
the center of mass mode $\bar x(t) = \frac{1}{N} \sum_i x_i(t)$ undergoes diffusion,
i.e. with ${\rm Var} \bar x(t) = \frac{2 T}{N} t$ when started from $\bar x(t)=0$ at $t=0$. 
This motion is subdominant in the limit studied here.
\\

{\bf Remark}. This stationary solution has the same form as the well known soliton of the focusing non-linear Schr\"odinger equation (NLSE). The NLSE also describes a semi-classical limit of the delta Bose gas. 
\\

{\it Convergence to the stationary state: droplet initial condition}. Let us study the dynamics
with initially all particles at $x=0$, hence $r_0(x)= \frac{1}{2} {\rm sgn}(x)$. Inserting into \eqref{rz}, \eqref{Z} 
or \eqref{soluCH} we obtain
\be \label{soludrop} 
r(x,t) = \frac{1}{\bar \gamma} \partial_x \log \left( e^{\frac{\bar \gamma}{2} x} (1 + {\rm erf}(\frac{x+ \bar \gamma t}{2 \sqrt{t}}) ) + e^{- \frac{\bar \gamma}{2} x} (1 - {\rm erf}( \frac{x- \bar \gamma t}{2 \sqrt{t}} ) ) 
\right) 
\ee 
In the large time limit, it converges to $r_{\rm stat} (x)$ given by \eqref{stat2} as
\be 
r(x,t) \simeq r_{\rm stat} (x) - \frac{e^{- \frac{\bar \gamma^2}{4} t }}{\bar \gamma^2 \sqrt{\pi t}} \partial_x \left( \frac{e^{-\frac{x^2}{4 t}}}{\cosh(\frac{\bar \gamma x}{2})} \right)
\ee 
This decay rate is compatible with the one predicted by the Lieb-Liniger model, 
$\simeq \frac{N^2 \bar c^2}{4}= \frac{\bar \gamma^2}{4}$ at large $N$, as expected
since the initial condition is very localized. 
\\

{\it Convergence to the stationary state: slow decaying solutions}.
Consider now the following initial condition with two packets of particles centered at positions $x_1$ and $x_2$
(in the remainder of this subsection we set $T=1$ for notational simplicity)
\bea
&& \int^w dx' r_0(x')  = \frac{1}{\bar \gamma}   \log[ \cosh[\frac{p_1 \bar \gamma}{2} (w -x_1)]  + \frac{1}{\bar \gamma}   \log[ \cosh[\frac{p_2 \bar \gamma}{2} (w -x_2)]] \quad , \quad p_1+p_2= 1 \label{2packets} \\
&& r_0(x) =  \frac{p_1}{2}  \tanh ( \frac{p_1 \bar \gamma}{2} (x-x_1) ) + \frac{p_2}{2} \tanh ( \frac{p_2 \bar \gamma}{2} (x-x_2) ) 
\eea
which has the proper boundary conditions $r_0(x)= \pm \frac{1}{2}$ at $x \to \pm \infty$.
If the packets are well separated one can consider that they contain respectively the fractions $p_1$ and $p_2$ 
of particles with $p_1+p_2=1$. Inserting \eqref{2packets}  into 
\eqref{Z} and \eqref{soluCH} one finds
\be 
r(x,t)= \frac{1}{\bar \gamma} \partial_x 
\log\left( \cosh( \frac{\bar \gamma}{2} (x - x_s) ) + e^{- p_1 p_2 \bar \gamma^2 t} 
\cosh( \frac{\bar \gamma}{2} ((p_1-p_2) x - (p_1 x_1-p_2 x_2)) )
\right) 
\ee 
which converges at large time to the stationary solution \eqref{stat2}, i.e to a single shock
containing all the particles at position
\be 
x_s = p_1 x_1 + p_2 x_2 
\ee 
At finite time this solution describes two coalescing shocks. 
Note that the decay rate $1/t_2$ is 
\be 
1/t _2= p_1 (1-p_1) \bar c^2 N^2 \leq \frac{1}{4} \bar c^2 N^2
\ee 
i.e. the decay is slower than the one predicted by the gap in the LL bosonic spectrum between 
the ground state and the one particle excitation. As explained in the text, this is because the
spatial decay of the initial density, here $\rho_0(x) \sim e^{- \min(p_1,p_2) N \bar c |x|}$ at large $x$ is 
slower than the one of $\Psi_0(\vec x=(x,0,\dots,0)) \sim e^{-  \frac{N-1}{2} \bar c |x|}$. 
\\

When the packets are well separated (as compared to their widths), one can describe their finite time dynamics 
by considering piecewise superpositions of elementary solutions interpolating 
between $a_j-\frac{p_j}{2}$ and $a_j+\frac{p_j}{2}$ of the form
\bea
r(x,t) = a_j + \frac{p_j}{2} \tanh( \frac{p_j}{2} \bar \gamma (x - x_j + 2 a_j  \gamma t) )
\eea 
with $a_j= - \frac{1}{2} + p_1+ \dots + p_{j-1}$. The center of each packet then moves balistically as
$x_s=x_j -2 a_j \gamma t$. This solution is a good approximation until the packets get closer together. 


\subsubsection{Dynamics for repulsive interactions $\gamma>0$}

To obtain some insight into the Burgers dynamics for the
expanding gas (repulsive case) we can either study 
the solution \eqref{soludrop} setting $\bar \gamma = - \gamma$ with
$\gamma>0$, or study again the case of the square initial condition 
$\rho(x,0)=\frac{1}{2\ell} \theta(\ell-|x|)$ with $a>0$, which was solved in \eqref{square1}
and \eqref{solurep} in the fixed $\bar c$ large $N$ limit (inviscid limit). Inserting 
into \eqref{soluCH} we obtain
\bea
&& r(x,t)= - \frac{1}{\gamma} \partial_x \log \left( f(x,t) + f(-x,t) \right) \\
&& f(x,t)= \frac{e^{-\frac{\gamma x^2}{4 ({\ell}+\gamma
   t)}}}{\sqrt{1+\frac{\gamma t}{{\ell}}}} 
 \text{erf}\left(\frac{{\ell}+\gamma
   t+x}{2 \sqrt{t} \sqrt{1+\frac{\gamma  t}{{\ell}}}}\right)
   +e^{\frac{1}{4}
   \gamma ({\ell}+\gamma t+2 x)}
   \text{erfc}\left(\frac{{\ell}+\gamma t+x}{2
   \sqrt{t}}\right)
\eea 
These solutions look like \eqref{solurep} but with a smoothed boundary.

\subsection{Harmonic potential, repulsive interactions} 

Consider the harmonic well $V(x)= \mu \frac{x^2}{2}$ with $\mu>0$.
In the simplest case of no interactions 
$\gamma=0$ the stationary solution is simply
\be
r_{\rm stat} (x) = \frac{1}{2} {\rm Erf}( \frac{x \sqrt{\mu}}{\sqrt{2 T}} ) \quad , \quad \rho_{\rm stat} (x)= \frac{\sqrt{\mu}}{\sqrt{2 \pi T} } 
e^{ - \frac{\mu x^2}{2 T}} \label{gamma0} 
\ee
\\

Consider now repulsive interactions $\gamma>0$. Defining 
$r(x,t) = \frac{\mu}{2 \gamma} x + \tilde r(x,t) $ (so that $\tilde r(x,t)=0$ in the stationary state in the absence of the second
derivative) one can rewrite \eqref{eqrho23} as
\bea \label{eqR}
\partial_t \tilde r =  T \partial_x^2 \tilde r  - 2  \gamma \tilde r \partial_x \tilde r -  \mu \tilde r
\eea 
which is the Burgers equation with friction $\mu$ for $\tilde r(x,t)$. 
\\

Let us search for the stationary solution $\partial_t \tilde r=0$ of \eqref{eqR}, which we call $\tilde r_{\rm stat} $. It is easy to see that 
$\tilde r_{\rm stat} $ takes the scaling form $\tilde r_{\rm stat} (x)= - \frac{\sqrt{\mu T}}{2 \gamma} R(x \sqrt{\frac{\mu}{T}})$,
hence $r_{\rm stat} (x)$ is of the form 
\be
r_{\rm stat} (x)= \frac{\sqrt{\mu T}}{2 \gamma} \hat r_g(x \sqrt{\frac{\mu}{T}}) \quad , \quad 
\hat r_g(y) = y - R_g(y)  \quad , \quad 
\rho_{\rm stat} (x)= \frac{\mu}{2 \gamma} \hat \rho_g(x \sqrt{\frac{\mu}{T}}) \quad , \quad \hat \rho_g(y)= 1 - R_g'(y)  
\ee 
where we have introduced the dimensionless parameter
\be 
g = \frac{\gamma}{\sqrt{\mu T}}
\ee 
The scaling functions form a one parameter family depending on the parameter $g$. 
The function $R(y)=R_g(y)$ is a dimensionless function solution of 
\bea \label{tobesolved} 
0 = R''(y)  + R(y) R'(y) -  R(y)
\eea 
which amounts to set $\mu=1$, $\gamma=1/2$ and $T=1$. 
Since $r(\pm \infty)= \pm \frac{1}{2}$, hence $R_g(y)$ must satisfy for $y \to \pm \infty$
\be \label{condR} 
R(y) \simeq y \mp g 
\ee 
and we expect $R_g(y)$ to be an odd function. The condition \eqref{condR} can be
equivalently written as
\be \label{condR2} 
\int_{-\infty}^{+\infty} dy (1-R'(y)) = 2 g 
\ee 
which will be useful below. 

The equation \eqref{tobesolved}  is an autonomous equation, $y$ does not appear. The usual method is to write
$R'(y) = w(R(y))$ for some function $w(R)$ of $R$. Then $R''(y)= w(R(y)) w'(R(y))$
and the equation \eqref{tobesolved} becomes an equation for $w(R)$ where now $R$ is the variable
\be
w w' =  R (1 - w) \quad \Rightarrow \quad \frac{w dw}{1-w} = R dR
\ee 
This can be integrated as
\be \label{eq10} 
w + \log( 1 - w ) = - \frac{1}{2} R^2 - b
\ee 
where $b$ is a constant as yet undetermined. Clearly $y \to \pm \infty$ corresponds to $R \to \pm \infty$
and to $w \to 1^-$. 

There are several equivalent ways to write the solution. The first one is to solve \eqref{eq10} 
for $w(R)$ as 
\be
w = 1 + W( - e^{-b-1 - \frac{R^2}{2}}) 
\ee 
where here $W(z)=\sum_{n \geq 1} \frac{(-n)^{n-1}}{n!} z^n$ is the first branch of the Lambert
function solution of $z=W e^W$, with $z \in [e^{-1} , +\infty[$ and $W(e^{-1})=-1$. 
The condition \eqref{condR2} implies that 
\be
\int_{-\infty}^{+\infty} (1-w) dx = \int_{-\infty}^{+\infty} \frac{1-w(R)}{w(R)}  dR = 2 g
\ee 
which allows to determine the constant $b=b_g>0$ as a function of $g$ as the solution of 
\be \label{condb} 
\int_{0}^{+\infty} du \, \left( \frac{1}{1 + W( - e^{-b-1 - \frac{u^2}{2}})} - 1 \right) = g
\ee 
Finally since $\frac{dy}{dR}= \frac{1}{w(R)}$ one obtains that $R(y)=R_g(y)$ is determined by inverting from 
\be \label{xx} 
\int_{0}^R  \frac{du}{1 + W( - e^{-b-1 - \frac{u^2}{2}})}= y 
\ee 
with $b=b_g$.
The scaling function of the density is then determined parametrically by eliminating $R$
between \eqref{xx} and
\be
\hat \rho(y) = - W(- e^{-b-1 - \frac{R^2}{2}})
\ee 
Thus the density depends on the parameter $b=b_g$ which depends on $g$ 
via \eqref{condb}. 

In the limit $g \ll 1$ one has $b \to +\infty$ and one can expand $W(z)$ at small $z$.
From \eqref{condb} one obtains 
\be
z = e^{-b-1} = \sqrt{\frac{2}{\pi }} g-\frac{2 \sqrt{2} g^2}{\pi
   }+\frac{\left(8 \sqrt{2}-3 \sqrt{6}\right) g^3}{\pi
   ^{3/2}}
   +O\left(g^5\right)
\ee 
Next one obtains
\bea 
R_g(y) = y-g \, 
   \text{erf}\left(\frac{y}{\sqrt{2}}\right)+\frac{g^2
   \left(\left(\sqrt{2} e^{-\frac{y^2}{2}}+2\right)
   \text{erf}\left(\frac{y}{\sqrt{2}}\right)-2
   \text{erf}(y)\right)}{\sqrt{\pi
   }}+O\left(g^3\right)
\eea 
and finally
\bea 
\hat \rho(y)= \sqrt{\frac{2}{\pi }} g e^{-\frac{y^2}{2}}+\frac{g^2
   e^{-y^2} \left(\sqrt{2 \pi } e^{\frac{y^2}{2}} y
   \text{erf}\left(\frac{y}{\sqrt{2}}\right)-2
   \sqrt{2} e^{\frac{y^2}{2}}+2\right)}{\pi
   }+O\left(g^3\right)
\eea 
The leading term recovers the result \eqref{gamma0} for $\gamma=0$ 

In the other limit $g \gg 1$ one finds that in the
variable $u=y/g$ the density goes to a square function
\be 
\hat \rho(y) \simeq \theta(1 - \frac{|y|}{g}) \quad ,\quad \rho(x) 
\simeq \frac{\mu}{2 \gamma} \theta(1 - \frac{|x| \mu}{\gamma})  
\ee 
Since in that limit $b \to 0$, one can check that indeed 
$\hat \rho(0) \to 1$ since
\be
W(- e^{-1-b}) = -1 + \sqrt{2 b} - \frac{2}{3} b + \dots
\ee 
\\

{\it Alternative method}. To solve the equation one can write instead, starting from \eqref{eq10}
\be
R(y) = F(w) \quad , \quad F(w) = \pm \sqrt{- 2 b - 2 (w + \log( 1 - w ) )}
\ee
From the definition $w = R'(x)$ one has 
\be
dx = \frac{dR}{w} = \frac{F'(w) dw}{w} 
\ee 
This leads to the solution in parametric form 
\bea
&& R = F(w) = \pm \sqrt{-2 b - 2 (w + \log( 1 - w ) )} = \pm \sqrt{2} \sqrt{-1-b+e^z-z} \\
&&  y = \int^w \frac{F'(u) du}{u} = \pm \int^w \frac{du}{\sqrt{2} (1-u) \sqrt{-b-u-\log (1-u)}}
= \pm \int_{z} \frac{dz}{\sqrt{2} \sqrt{-b-1+e^z-z}} 
\eea 
defining $1-w=e^z$. 
 \\
 
 {\bf Remark}. Similar looking equations appear in a problem of self-gravitating gas \cite{Kumar2017}, but with somewhat 
 different interactions, which are linear and attractive at small scale and quadratic and repulsive 
 at large scales (and there is no external quadratic well). 

\section{Coulomb gas}

Here we discuss the Coulomb gas formulation, which also allows to determine the stationary state
in the large $N$ limit. 

The (exact) Dean-Kawasaki equation \eqref{eqrho1} for the RD system in terms of the density field $\tilde \rho(x,t)= \sum_i \delta(x-x_i(t)))$ normalized to $N$,
and restoring the temperature $T$, reads
\be \label{eqrho1app}
 \partial_t \tilde \rho(x,t)  =   T \partial_x^2 \tilde  \rho(x,t)  + \partial_x [ \sqrt{ 2 T \tilde  \rho(x,t) } \eta(x,t) ]  +  \partial_x [ V'(x) \tilde  \rho(x,t) +
 \bar c \tilde  \rho(x,t) \int dy \tilde  \rho(y,t) {\rm sgn}(x-y) )] 
\ee 
where $\eta(x,t)$ is a unit space time white noise. Following \cite{Dean,DeanPrivate} it can be rewritten as an explicit equilibrium dynamics 
\be \label{eqdyn} 
 \partial_t \tilde \rho(x,t) = - \int dy R(x,y;\tilde \rho(.,t)) \frac{\delta H[\tilde \rho(.,t)]}{\delta \tilde \rho(y,t)} + \xi(x,t) \quad , \quad  \langle \xi(x,t) \xi(y,t) \rangle = 2 T R(x,y;\tilde \rho(.,t)) \delta(t-t') 
\ee
where 
\be 
R(x,y;\tilde \rho)= \partial_x \partial_y ( \sqrt{\tilde \rho(x) \tilde \rho(y)} \delta(x-y)) = \partial_x ( \tilde \rho(x,t) \partial_y \delta(x-y)) 
\ee
and the energy functional is 
\be
H[\tilde \rho]= \int dx V(x)  \tilde \rho(x) + \frac{\bar c}{2} \int  dx dx' |x-x'| \tilde \rho(x) \tilde \rho(x')  
+ T \int dx \tilde \rho(x) \log \tilde \rho(x) 
\ee 
The last term is the so-called entropy term. Indeed one has
\be 
\frac{\delta H[\tilde \rho]}{\delta \tilde \rho(y)} = V(y) + \bar c \int dx' |y-x'| \tilde \rho(x') + T (1 + \log \tilde \rho(y)) 
\ee 
Using that for any function $f(y)$, upon integration by part, 
\be 
- \int dy R(x,y;\tilde \rho) f(y) = - \int dy \partial_x ( \tilde \rho(x) \partial_y \delta(x-y)) f(y) 
= \partial_x ( \tilde \rho(x) f'(x) )
\ee 
one sees that \eqref{eqdyn} is equivalent to \eqref{eqrho1app}. 
\\

Thus the stochastic dynamics of the system satisfies detailed balance,
and generically converges to the equilibrium measure
\be \label{Gibbs} 
{\cal P}_{\rm stat}[\tilde \rho] \propto \exp\left(  - \frac{H[\tilde \rho]}{T} \right) 
\ee 
whenever the latter is well defined. Note that it does not say how fast this convergence
holds. It thus describes the limit where time is taken to infinity first. 

Let us now consider the two large $N$ limits studied here.

\subsection{Large $N$ at fixed $c=-\bar c$, $V(x)=N \tilde V(x)$} 

Let define, as in the text, $\rho(x,t) = \frac{1}{N} \tilde \rho(x,t) = \frac{1}{N} \sum_i \delta(x-x_i(t)))$
the density normalized to unity. In this regime we consider large $N$ at fixed $c=-\bar c$, and 
scale the potential as $V(x)=N \tilde V(x)$ with fixed $\tilde V(x)$. The energy becomes
\be 
H[\tilde \rho] = N^2 {\cal E}[\rho] + O(N) \quad , \quad {\cal E}[\rho]= - \frac{c}{2} \int  dx dx' |x-x'| \rho(x) \rho(x')  
+ \int dx \tilde V(x)  \rho(x) \label{Erho} 
\ee 
where we can neglect the entropy term in that regime since it is $O(N)$. The equilibrium measure
$P_{\rm stat} \propto e^{- N^2 {\cal E}[\rho]/T}$, is dominated by the energy minimum.
To minimize under 
the constraint $\int dx \rho(x)=1$ one adds the Lagrange multiplier term $\nu (\int dx \rho(x)-1)$ to ${\cal E}[\rho]$.
The saddle point equation gives 
\be \label{firsteq} 
 \frac{\delta {\cal E}}{\delta \rho(x)} =  - c  \int  dx' |x-x'| \rho(x')  + \tilde V(x) + \nu = 0 
\ee 
for any $x$ in the support of $\rho(x)$. Taking a derivative we find that either $\rho(x)=0$ or 
\bea
 \partial_x \frac{\delta {\cal E}}{\delta \rho(x)} = - c  \int  dx'  {\rm sgn}(x-x') \rho(x') + \tilde V'(x) = - 2 c \, r(x) + \tilde V'(x)  = 0 
\eea
where we used that $\int  dx'  {\rm sgn}(x-x') \rho(x') = 2 r(x)$. Hence we find 
that for any point $x$ inside the support of the density one has $r(x)=\tilde V'(x)/(2 c)$ and 
$\rho(x)=\tilde V''(x)/(2 c)$ (and, since $-1/2 \leq r(x) \leq 1/2$, any point such that $\tilde V'(x) > c$ 
cannot belong to the support). This coincides with the result \eqref{aga} which was obtained
by considering the large time limit for $c>0$ (the limit $N \to \infty$ being taken first). 
There we gave a complete discussion
of the determination of the support for various types of potentials $\tilde V(x)$. 
We would like to know how it compares with the Coulomb gas formulation. 
Let us note that the minimum energy verifies,
using \eqref{firsteq}
\be 
{\cal E}_{\rm min} = \int dx \tilde V(x)  \rho(x) - \frac{1}{2} \nu
\ee 

Let us check whether \eqref{firsteq} is verified. Suppose that the support is a union of (ordered)
intervals $[a_i,b_i]$. Each interval contributes to the first term in the RHS of \eqref{firsteq} as
\bea
&& - c  \int_{a_i}^{b_i}  dx' |x-x'| \rho(x') = - \frac{1}{2} \int_{a_i}^{b_i}  dx' |x-x'|  \tilde V''(x') 
= \frac{1}{2} \int_{a_i}^{b_i}  dx' {\rm sgn}(x'-x)  \tilde V'(x')  - \frac{1}{2}  [ |x-x'|  \tilde V'(x') ]_{a_i}^{b_i} 
\\
&& =  - \int_{a_i}^{b_i}  dx' \delta(x-x') \tilde V(x')  + \frac{1}{2} [{\rm sgn}(x'-x)  \tilde V(x')]_{a_i}^{b_i} 
- \frac{1}{2}  [ |x-x'|  \tilde V'(x') ]_{a_i}^{b_i} 
\eea
If $x$ belongs to one of the interval, then \eqref{firsteq} holds provided the following sum is a constant in $x$
\be \label{cond10} 
\sum_i \frac{1}{2} [{\rm sgn}(x'-x)  \tilde V(x')]_{a_i}^{b_i} 
- \frac{1}{2}  \sum_i  [ |x-x'|  \tilde V'(x') ]_{a_i}^{b_i} = - \nu 
\ee 
Let us examine several cases
\\

{\it Single interval}. Consider first the case where the support is a single interval $[a_1,b_1]$, 
which is the case for convex potentials, as in Fig. \ref{Fig:Rep1}
(with $a_1=x_e^-$ and $b_1=x_e^+$). In that case we know that $\tilde V'(a_1)= - c$
and $\tilde V'(b_1)=c$. The first term gives $\frac{1}{2} (\tilde V(b_i)+ \tilde V(a_i))$ and the second 
$- \frac{c}{2} (b_i - a_i)$, which are constants. 
Hence \eqref{firsteq} is verified with $\nu= - \frac{1}{2} (\tilde V(b_i)+ \tilde V(a_i)) + \frac{c}{2} (b_i - a_i)$ and the energy at the minimum is then 
\be 
{\cal E}_{\rm min} = \frac{1}{2 c} \int_{a_1}^{b_1} dx \tilde V(x)  \tilde V''(x)  
+  \frac{1}{4} (\tilde V(b_i)+ \tilde V(a_i)) - \frac{c}{4} (b_i - a_i)
\ee 
In the case of the quadratic potential one finds ${\cal E}_{\rm min} =- \frac{1}{6} c^2/\mu_0$. 
\\

{\it Two intervals}. Consider now the case where the support consists in two intervals, as is the case for the double well potential, 
as represented in Fig. \ref{Fig:Rep2} 
with $a_i=x_e^{i,-}$ and $b_i=x_e^{i,+}$. We have (see the figure)
\be \label{cond11} 
\tilde V'(a_1)= - c \quad , \quad \tilde V'(b_2)= +c \quad , \quad \tilde V'(a_2) = \tilde V'(b_1) 
\ee 
which implies the normalisation $\int dx \rho(x)= \frac{1}{2 c} ( \int_{a_1}^{b_1} \tilde V''(x) + \int_{a_2}^{b_2} \tilde V''(x) ) = 1$. We obtain the two terms in \eqref{cond10}
\be \label{bothsigns} 
- \nu_\pm := \frac{1}{2} ( \pm \tilde V(b_1)+ \tilde V(a_1) + \tilde V(b_2) \mp \tilde V(a_2) ) - \frac{1}{2} ( \pm b_1 \tilde V'(b_1) + a_1 \tilde V'(a_1) + b_2 \tilde V'(b_2)  \mp a_2 \tilde V'(a_2)) = - \nu 
\ee 
where the plus sign is for $x$ in the first interval and the $-$ sign in the second. Note that the
term linear in $x$ cancels using \eqref{cond11}. In the dynamics we found that 
the height of the plateau in the rank field, $r(b_1)= \frac{\tilde V'(b_1)}{2 c}=r(a_2)=\frac{\tilde V'(b_2)}{2 c}$,
was determined by the initial condition. Here we see that for both signs in \eqref{bothsigns} to give the same (i.e. compatible) result, we need in addition to \eqref{cond11}
\be
\tilde V(b_1) - b_1 \tilde V'(b_1) = \tilde V(a_2)- a_2 \tilde V'(a_2) 
\ee
These two conditions can also be written as
\be \label{cond12} 
\int_{a_2}^{b_1} dx \, \tilde V''(x) = 0 \quad , \quad \int_{a_2}^{b_1} dx \, x \tilde V''(x) = 0 
\ee
which determine the position of the plateau in the Coulomb gas. For instance for
a symmetric double well $\tilde V(x)= \frac{x^4}{12}-\frac{x^2}{2}$ we obtain that the 
plateau is at level $r(b_1)=r(a_2)=0$, i.e. $a_2=-b_1=\sqrt{3}$ (setting $c=1$) with
$V'(a_2)=V'(b_1)=0$. We have checked that this indeed realizes the minimum of the energy
${\cal E}(\rho)$ as compared to allowing different levels for the plateau as obtained in the dynamics.
\\

To conclude this subsection, the Coulomb gas gives the equilibrium state of the system, i.e.
the infinite time limit is taken first, to obtain the Gibbs measure \eqref{Gibbs}, and only in a second
stage the large $N$ limit is performed on this equilibrium state. In the dynamics of Section 
\ref{sec:fixed} we have performed large $N$ first and neglected the noise. In cases such as double well potentials, convergence to the true equilibrium requires some noise to cross barriers. 
Hence at large $N$ there are two time scales in the dynamics. 
First a rapid convergence to a metastable state as in Fig. \ref{Fig:Rep2}, where the 
number of particles in each well is determined by the initial condition. This stage is
captured by the dynamics of Section 
\ref{sec:fixed}. Next, on much
larger time scales, particles undergo barrier crossing, so that the "chemical potentials" 
$\nu_\pm$ in each well become identical, leading to the equilibrium plateau \eqref{cond12}. 
It would be interesting to study that slow dynamics. 
\\

{\bf Remark}. It is interesting to note that one can rewrite the energy \eqref{Erho} using integrations by part in terms of the rank field as
\be 
{\cal E}[\rho]= c  \int dx (r(x)- \frac{1}{2} {\rm sgn}(x))^2 + \int dx \, (c \, {\rm sgn}(x) -  \tilde V'(x) ) (r(x)- \frac{1}{2} {\rm sgn}(x)) 
\ee 
\\

{\it Attractive case $c=-\bar c<0$}. In that case it is easy to see that the electrostatic energy (which is positive) is minimum by itself
(and vanishes) for $\rho(x) = \delta(x-x_s)$, which correspond to a single shock at position $x_s$ containing 
a unit fraction of the particles. In the absence of a potential $x_s$ is arbitrary. 
In the presence of an external potential, the potential energy will be minimized (also by itself) 
if $x_s$ is chosen at the minimum of this potential (assuming that there is one unique such minimum).

\subsection{Large $N$ at fixed $\gamma= N c$} 

In that limit one has (up to a constant)
\be 
H[\tilde \rho] = N {\cal E}_2[\rho]  \quad , \quad {\cal E}_2[\rho]= - \frac{\gamma}{2} \int  dx dx' |x-x'| \rho(x) \rho(x')  
+ \int dx V(x)  \rho(x) + T \int dx \tilde \rho(x) \log \tilde \rho(x) 
\ee 
The problem still maps to a Coulomb gas, with a different scaling for
the equilibrium measure $P_{\rm stat} \propto e^{- N {\cal E}_2[\rho]/T}$. It is still
dominated by the minimum energy configuration, but this time the entropy term becomes important.

The saddle point equation now gives (absorbing constants in $\nu$) 
\be \label{firsteq} 
 \frac{\delta {\cal E}}{\delta \rho(x)} =  - \gamma  \int  dx' |x-x'| \rho(x')  + V(x) + T \log \rho(x) + \nu = 0 
\ee 
for any $x$ in the support of $\rho(x)$. Taking a derivative we find that either $\rho(x)=0$ or 
\bea
 \partial_x \frac{\delta {\cal E}}{\delta \rho(x)} = - 2 \gamma \, r(x) + \tilde V'(x)  + T \frac{\rho'(x)}{\rho(x)}  = 0 
\eea
where we used again that $\int  dx'  {\rm sgn}(x-x') \rho(x') = 2 r(x)$. This equation can be rewritten as
\be
0 = T \partial_x^2 r - 2 \gamma r \partial_x r + V'(x) \partial_x r 
\ee 
which is exactly the stationary equation associated to the Burgers equation 
in an external potential \eqref{eqrho22} obtained in the text. It was solved explicitly in the previous
section (i) for repulsive interactions $\gamma>0$ in the case of the quadratic potential $V(x)=\frac{\mu}{2} x^2$ 
(ii) for attractive interactions $\gamma<0$ and $V(x)=0$.

\end{widetext}


\begin{thebibliography}{}
\bibitem{footnotecdm} The center of mass 
decouples from the relative coordinates
and undergoes free diffusion: its invariant measure is uniform.
\bibitem{footnoteC} The constant $C$ can be chosen 
arbitrarily (we use unnormalized wave functions)
and does not enter the final results.
\bibitem{footnote2} For $V(x)=0$ and $\bar c<0$ it cannot be 
the ground state on the infinite line since its normalization 
diverges exponentially. Indeed, the ground state, e.g. on
a large circle of perimeter $L \to +\infty$ has a quite different structure.
\bibitem{footnote1}
Note that for $\tilde \rho(x,t) = \sum_i \delta(x-x_i(t))$, normalized to $N$, the same
equation holds with no factors of $N$.
\end{thebibliography}
\end{document}